\documentclass[a4paper,oneside,final,notitlepage,onecolumn]{article}
\usepackage{amsfonts}
\usepackage{epsf}
\usepackage{graphicx}
\usepackage{amssymb}

\usepackage{t1enc}
\DeclareFontFamily{OT1}{rsfs}{} \DeclareFontShape{OT1}{rsfs}{m}{n}{
<-7> rsfs5 <7-10> rsfs7 <10-> rsfs10}{}
\DeclareMathAlphabet{\mycal}{OT1}{rsfs}{m}{n}

\def\scri{{\mycal I}}
\def\scrip{\scri^{+}}%

\def\sc{{\hskip 3.5pt {{}^{{}^{{}_{{}_{\bowtie}}}}} \kern -8.pt{}}}  
\def\SC{{\hskip 3.5pt {{}^{{}^{{}^{{}_{{}_{\bowtie}}}}}} \kern -10.5pt{}}}

\begin{document}

\newtheorem{theorem}{Theorem}[section]
\newtheorem{lemma}{Lemma}[section]
\newtheorem{proposition}{Proposition}[section]
\newtheorem{corollary}{Corollary}[section]
\newtheorem{conjecture}{Conjecture}[section]
\newtheorem{example}{Example}[section]
\newtheorem{definition}{Definition}[section]
\newtheorem{remark}{Remark}[section]
\newtheorem{exercise}{Exercise}[section]
\newtheorem{axiom}{Axiom}[section]
\renewcommand{\theequation}{\thesection.\arabic{equation}} 

\author{ Istv\'an R\'{a}cz$^{1,2}$\thanks{%
~Research Fellow of the Japan
Society for the Promotion of Science, email:
iracz@yukawa.kyoto-u.ac.jp}  \\ 
$^1$Yukawa Institute for
Theoretical Physics\\ Kyoto University, Kyoto 606-01, Japan \\ 
$^2$MTA
KFKI, R\'eszecske- \'es Magfizikai Kutat\'oint\'ezet, \\ H-1121
Budapest, Konkoly Thege Mikl\'os \'ut 29-33. \\Hungary
}
\title{Stationary Black Holes as Holographs}  

\maketitle

\begin{abstract}

Smooth spacetimes possessing a (global) one-parameter group of
isometries and an associated Killing horizon in Einstein's theory of
gravity are investigated.  No assumption concerning the asymptotic
structure is made, thereby, the selected spacetimes may be considered
as generic distorted stationary black holes.  First, spacetimes of
arbitrary dimension, $n\geq 3$, with matter satisfying the dominant
energy condition and allowing non-zero cosmological constant are
investigated.  In this part complete characterisation of the topology
of the event horizon of ``distorted'' black holes is given.  It is
shown that the topology of the event horizon of ``distorted'' black
holes is allowed to possess a much larger variety than that of the
isolated black hole configurations.  In the second part,
$4$-dimensional (non-degenerate) electrovac distorted black hole
spacetimes are considered.  It is shown that the spacetime geometry
and the electromagnetic field are uniquely determined in the black
hole region once the geometry of the bifurcation surface and one of
the electromagnetic potentials are specified there. Conditions
guaranteeing the same type of determinacy, in a neighbourhood of the
event horizon, on the {\it  domain of outer communication} side are
also investigated. In particular, they are  shown to be satisfied in
the analytic case.

\end{abstract}

\section{Introduction}
\setcounter{equation}{0}

The most significant part of our insight into black holes originates
from the study of the well-known Schwarzschild  and Kerr black
holes. According to the powerful black hole uniqueness theorems (see
e.g. Refs. \cite{carter1,carter2,bunting,HE}) they are known to be
{\it the only} asymptotically flat stationary black hole solutions of
the vacuum Einstein equations. On one hand, the requirement of
asymptotically flatness is natural to be imposed whenever one is
interested in the properties of black holes which are completely
isolated in space. On the other hand, it is also of great importance
to know how these isolated black holes might be distorted by external
matter or other mass distributions as it has to happen in all
physically realistic situations. Therefore, it is of crucial interest
to determine all the possible distorted black hole solutions and also
to provide a clear characterisation of them.  In this paper we find
all stationary distorted black holes in Einstein theory of gravity,
moreover, some of their generic properties are also investigated.

\medskip

To understand our motivations in setting up the mathematical framework
we shall apply let us recall first that a key result in the black
holes uniqueness proofs is the so-called black hole rigidity theorem
of Hawking \cite{hawk1, HE}, which asserts that, in Einstein's theory
of gravity, the event horizon of an {\it analytic} stationary
asymptotically flat electrovac black hole spacetime is  necessarily a
Killing horizon, i.e., the spacetime must possess a Killing field
(possibly distinct from the asymptotically stationary Killing field)
which is normal to the event horizon. A consequence of this result is
that in an asymptotically flat stationary (non-static) black hole
spacetime there exists an additional axial Killing vector field, i.e.,
a stationary black hole spacetime is either static or stationary
axisymmetric. With the help of these results the uniqueness of
electrovac black holes can be proved by showing the uniqueness of
solutions to an elliptic boundary value problem
\cite{israel1,israel2,carter1,carter2,mazur,bunting} where the
elliptic equations are derived from the Einstein's equations on the
space (or on a suitable factor space) of the ``$t=const$''
hypersurfaces while the boundary values are specified at locations
corresponding to the bifurcation surface of the event horizon and at
spacelike infinity \cite{bunting}. It was (implicitly) assumed in the
corresponding  arguments that in the non-degenerate case the black
hole event horizon is a  bifurcate type  Killing horizon. The validity
of this assumption has been justified by a series of papers for the
smooth geometrical setting either for generic metric theories of
gravity \cite{rw1,rw2} or  in general relativity with the inclusion of
various matter fields \cite{frw,r1}.

\medskip

Recently, the above recalled classic result of Hawking has been
generalised to the case of higher dimensional stationary black hole
spacetimes by Hollands, Ishibashi and Wald \cite{hiw}. In particular,
it was justified by them that the event horizon of an asymptotically
flat analytic stationary (non-degenerate) black hole must be a Killing
horizon even in the case of higher dimensional spacetimes. Therefore,
there seems to be no loss of generality in representing generic
stationary distorted black holes by spacetimes possessing a Killing
horizon which will be done in this paper.  It seems also to be
reasonable that in a framework suitable to investigate the properties
of  distorted stationary black holes  the assumption of asymptotic
flatness should be relaxed.  Therefore, no restriction concerning the
asymptotic structure will be imposed which, in particular, means we
shall {\it not} restrict our considerations to asymptotically flat or
asymptotically (locally) anti-de-Sitter spacetimes.

\medskip

Obviously, the black hole uniqueness results exclude the possibility
of having, e.g. asymptotically flat stationary vacuum black hole
solutions  other than the members of the Kerr-family. Nevertheless,
the  more generic class of spacetimes we are dealing with contains all
of those configurations which may be relevant in the context of
``distorted black holes''. Recall that all the static axially
symmetric vacuum ``distorted black hole'' configurations are known and
their properties have been studied extensively---for more details
concerning these spacetimes see,  e.g., Refs.
\cite{israel3,szekeres,PE,GH,CH,ME,FR1,HO,FR2,FK,Tom,FR3}. A limited
subclass of electrovac static axisymmetric distorted black hole
spacetimes has also been described and investigated in
\cite{FK}. Originally the static distorted black hole solutions were
considered to be relevant  only locally by representing a black hole
solution yielded by the distortion of the Schwarzschild solution by
certain external mass distributions. Nevertheless, later it was also
noticed that the generic distorted static black hole spacetimes  might
also play an important role in context of four (or higher) dimensional
black holes whenever one (or some) of the spacelike dimensions is (or are)
compactified (see e.g. Refs.\,\cite{ME,HO,FR2}).

\medskip

In this paper  spacetimes admitting a global one-parameter family of
isometry actions and an associated Killing horizon will be
investigated in Einstein's theory of gravity. In the first part the
topology of the Killing (or event) horizon of these spacetimes will be
considered. In this part the spacetime dimension will be assumed to be
arbitrary ($n\geq 3$), moreover, concerning the matter content only
the dominant energy condition will be required to be satisfied and we
shall also allow the inclusion of non-zero cosmological constant. It
is shown then that the topology of the Killing (or event) horizon is
much less restricted than that of the isolated black hole
configurations. In particular, in case of a $4$-dimensional spacetime
with non-positive cosmological constant\footnote{%
Notice that positive cosmological constant---in our metric signature
which is $(+,-,\dots,-)$---is associated with anti-de Sitter type
configurations.}  whenever the smooth global cross-sections of the
horizon are ``convex on the average''---in the sense that the null
geodesic congruences transverse to the event horizon are guaranteed to
be non-contracting on the  average towards the ``domain of outer
communication''---they possess the topology of either a sphere or a
torus, while the global cross-sections may be of higher genus compact
orientable $2$-surfaces whenever the cosmological constant is
positive.

In the second part, attention will be restricted to the case of
$4$-dimensional electrovac distorted black hole spacetimes with
non-degenerate Killing horizon. By making use of a combination of the
Newman-Penrose formalism \cite{newman:penrose} and that of the null
characteristic initial value formulation of Friedrich \cite{friedrich}
the true physical degrees of freedom of the selected class of
stationary electrovac black holes will be explored. It is shown that
the geometry and the electromagnetic field in the black hole region of
each of these spacetimes are uniquely determined by the specification
of the  $2$-metric of the cross-sections of the event horizon, along
with that of one of the electromagnetic potentials there. Within the
Newman-Penrose formalism a reduced set of quasilinear first order
partial differential equations (PDEs) are derived which determine the
spacetime metric and the electromagnetic field on both sides of the
event horizon. Then conditions guaranteeing the unique determinacy of
the investigated  distorted black hole spacetimes---in terms of the
$2$-metric of the bifurcation surface and that of one of the
electromagnetic potentials there---, in a neighbourhood of the event
horizon, on the {\it  domain of outer communication} side are also
investigated. In particular, it is shown that the relevant first order
quasilinear PDEs do possess unique solutions on domain of outer
communication side in the analytic case.  We would like to emphasise
that since the initial data---consisting of the $2$-metric of the
bifurcation surface and the relevant electromagnetic potential---is
completely free, thereby the associated spacetimes do not possess
any  spacetime symmetry besides the one associated with the Killing
horizon whence, in particular, they need not be axially symmetric
either.

\medskip

This paper is organised as follows: In Section\,\ref{stac} we specify
the class of black hole spacetimes to which our main results apply.
Section\,\ref{pre} starts by providing an introduction of the notions
of elementary spacetime regions and Gaussian null coordinates. The
following part, Section\,\ref{top}, is devoted to the study of the
topological properties of the selected spacetimes.  Then, in
Section\,\ref{pre2}, some details of the applied mathematical
techniques and the relevant results will be recalled. This part is
followed by an immediate application of the associated techniques to
the selected class of four dimensional vacuum stationary black hole
spacetimes.  Section\,\ref{key} is to summarise the main features of
the investigated black hole spacetimes, while in Section\,\ref{EM} the
consequences of the differences which will show up in the electrovac
case, with or without the inclusion of non-zero cosmological constant,
will be discussed briefly. Section\,\ref{con} contains our final
remarks and the addressing of some open issues.

\section{Preliminaries}\label{stac}
\setcounter{equation}{0}

Throughout this paper a spacetime $(M,g_{ab})$ is taken to be an
$n$-dimensional ($n\geq3$),   
smooth, paracompact, connected, orientable manifold $M$ endowed with a
smooth Lorentzian metric $g_{ab}$ of signature $(+,-,\dots,-)$. It is
assumed that $(M,g_{ab})$ is  time orientable and that a time
orientation has been chosen.

In the first part of this paper a spacetime $(M,g_{ab})$ will only be
assumed to satisfy the Einstein equations 
\begin{equation}\label{ein}
R_{ab}-\frac{1}{2} g_{ab} R+\widetilde\Lambda g_{ab}=8\pi T_{ab},
\end{equation}
where $\widetilde\Lambda$ stands for the cosmological constant, with
energy-momentum tensor, $T_{ab}$, satisfying the dominant energy
condition. Recall that the dominant  energy condition
is said to be satisfied if for all future directed timelike vector
$\xi^a$ the contraction ${T^a}_b\xi ^b$ is future
directed timelike or null vector \cite{wald}.

As indicated above in the second part $4$-dimensional electrovac
spacetimes will be considered. The electromagnetic field is assumed to
be  represented by a $2$-form field $F_{\,ab}$ which satisfies, in
addition to (\ref{ein}), the Maxwell equations
\begin{equation}\label{m1}
\nabla^aF_{\,ab}=0\ \ \ {\rm and}\ \ \ \nabla_{[a}{F}_{\,b c]}=0, 
\end{equation}
moreover, 
the energy momentum tensor, on the r.h.s. of  (\ref{ein}), is supposed
to take the form
\begin{equation}\label{max}
T_{ab}=-\frac{1}{4\pi}\left[ F_{\,ea}{F_{\,b}}^e -
\frac{1}{4} g_{ab} \left(F_{\,ef}F^{\,ef}\right) \right],
\end{equation}
which automatically satisfies the dominant
energy condition.

\medskip

Throughout this paper it is assumed that the spacetime $(M,g_{ab})$
admits a (global)  one-parameter group of isometries, $\chi_u$,
generated by a Killing vector field $\mathfrak{K}^a$. It will be also
required that $\chi_u$ admits a Killing horizon. Recall that a null
hypersurface $\mathcal{N}$ of $M$ is a Killing horizon, with respect
to $\chi_u$, whenever $\mathcal{N}$ is invariant under the action of
$\chi_u$, moreover, $\mathfrak{K}^a$ is null on
$\mathcal{N}$. We shall assume that $\mathfrak{K}^a$ is future
directed on $\mathcal{N}$, moreover, for simplicity, $\mathcal{N}$
will be assumed to be connected. Following \cite{rw1}, it will
also be required that
$\chi_u$ has no fixed point on $\mathcal{N}$, furthermore,
$\mathcal{N}$ is smooth admitting a smooth global cross-section
$\mathcal{Z}$. This, in particular, means that the orbits of $\chi_u$
are diffeomorphic to $\mathbb{R}$, moreover, each orbit of $\chi_u$
intersect $\mathcal{Z}$ precisely once, i.e., $\mathcal{N}$
necessarily possesses the product space structure $\mathbb{R}\times
\mathcal{Z}$. Finally, in order to restrict our attention to black
hole type configurations, we shall assume that $\mathcal{Z}$ is an
orientable $n-2$-dimensional compact manifold with no boundary.

\begin{definition} 
Hereafter, the spacetimes satisfying all the above generic conditions
will be referred to as spacetimes of class A, while the special
$4$-dimensional electrovac spacetimes will be referred to as spacetimes
of class B if the electromagnetic field, $F_{\,ab}$, is also invariant
under the  action of the one-parameter group of isometries, $\chi_u$.
\end{definition} 

Clearly, each spacetime of class B also belongs to the set of
spacetimes of class A.

\section{Gaussian null coordinates}\label{pre}
\setcounter{equation}{0}

Consider now a smooth spacetime $(M,g_{ab})$ of class A and   the
$1$-parameter family of smooth cross-sections
$\mathcal{Z}_u=\chi_u[\mathcal{Z}]$ of $\mathcal{N}$.  Choose then
$\mathfrak{L}^a$ to be the unique {\it future directed} null vector
field on $\mathcal{N}$ which is everywhere orthogonal to the
$2$-dimensional cross-sections $\mathcal{Z}_u$ and satisfies the
normalising condition $\mathfrak{L}^a\mathfrak{K}_a=1$ everywhere on
$\mathcal{N}$. Consider now the null geodesics starting at the points
of $\mathcal{N}$ with tangent $\mathfrak{L}^a$. Since $\mathcal{N}$
was assumed to be smooth, as well as,   the vector fields
$\mathfrak{K}^a$ and $\mathfrak{L}^a$ on $\mathcal{N}$ by construction
are also smooth, these geodesics do not intersect in a sufficiently
small open neighbourhood $\mathcal{O} \subset M$ of
$\mathcal{N}$. Such a neighbourhood $\mathcal{O}$ of $\mathcal{N}$
will be referred to as ``{\it elementary spacetime region}\,''. By
choosing $r$ to be the affine parameter along the null geodesics
starting at the points of $\mathcal{N}$ with tangent $\mathfrak{L}^a$
and synchronised so that $r=0$ on $\mathcal{N}$ we get a smooth real
function $r:\mathcal{O}\rightarrow \mathbb{R}$. The function
$u:\mathcal{N}\rightarrow \mathbb{R}$, which is smooth by
construction, can also be smoothly extended onto $\mathcal{O}$ by
requiring its value to be constant along the null geodesics with
tangent $\mathfrak{L}^a=\left(\partial/\partial r\right)^a$. Let us
denote also by $\mathfrak{K}^a$ the associated ``coordinate basis
field'', i.e., $\mathfrak{K}^a=\left(\partial/\partial u\right)^a$.

\medskip

Now, based on the smooth null hypersurface $\mathcal{N}$, and the
functions $u,r: {\mathcal{O}}\rightarrow \mathbb{R}$ already defined
on $\mathcal{O}$, Gaussian null coordinates
$\left(u,r,x^3,\dots,x^n\right)$ can be defined on suitable subsets
$\widetilde{\mathcal{O}}$ of $\mathcal{O}$, which comprise by
themselves ``elementary spacetime neighbourhoods'' of certain
subsections, $\widetilde{\mathcal{N}}$, of $\mathcal{N}$ as
follows. Let us choose, first, $\widetilde{\mathcal{Z}}$ to be a
connected open subset of the cross-section $\mathcal{Z}$ on
which local coordinates $\left(x^3,\dots,x^n\right)$ can be
defined. Choose, furthermore, $\widetilde{\mathcal{N}}$ be that subset
of $\mathcal{N}$ which is span by the null generators of $\mathcal{N}$
through the points of $\widetilde{\mathcal{Z}}$. Extend, then, the
functions $x^3,\dots,x^n$, first from $\widetilde{\mathcal{Z}}$ onto
$\widetilde{\mathcal{N}}$, and, second, from $\widetilde{\mathcal{N}}$
to $\widetilde{\mathcal{O}}$---consisting of exactly those points of
${\mathcal{O}}$ which can be achieved along the null geodesics
starting on $\widetilde{\mathcal{N}}$ with tangent $\mathfrak{L}^a$---so that
their values are kept to be constant, first along the generators of
$\widetilde{\mathcal{N}}$, second along the null geodesics with
tangent $\mathfrak{L}^a=\left(\partial/\partial r\right)^a$. 

Since by construction the vector field $\mathfrak{L}^a=\left( \partial
/\partial r\right) ^a$ is everywhere tangent to null geodesics we have
that $g_{rr}=0$ throughout $\widetilde{\mathcal{O}}$. Moreover, we
also have that the metric functions $g_{ru},g_{r3},\dots,g_{rn}$ are
independent of the $r$-coordinate, i.e.
$g_{ru}=1,g_{r3}=\dots=g_{rn}=0$ throughout $\widetilde{\mathcal{O}}$.
In addition, as a direct consequence of the above construction,
$g_{uu}$ and $g_{uA}$ vanish on $\widetilde{\mathcal{N}}$.  Hence,
within $\widetilde{\mathcal{O}}$, there exist smooth functions $f$ and
$h_A$, with $f\vert_{\widetilde{\mathcal{N}}}=(\partial
g_{uu}/\partial r)\mid _{r=0}$ and
$h_A\vert_{\widetilde{\mathcal{N}}}=(\partial g_{uA}/\partial
r)\mid_{r=0}$, so that the spacetime metric in
$\widetilde{\mathcal{O}}$ takes the form
\begin{equation}
\mathrm{d}s^2=r\cdot f\mathrm{d}u^2+2\mathrm{d}r\mathrm{d}u+2r\cdot
h_{A}\mathrm{d}u \mathrm{d}x^A+g_{AB}\mathrm{d}x^A\mathrm{d}x^B,
\label{le1}
\end{equation}
where $f$, $h_A$ and $g_{AB}$ are smooth functions of the coordinates
$r,x^3,\dots,x^n$ in $\widetilde{\mathcal{O}}$ such that $g_{AB}$ is a
negative definite $(n-2)\times (n-2)$ matrix, and the uppercase Latin
indices take the values $3,\dots,n$.

It is straightforward to see that, in the present case, the functions
$f$, $h_A$ and $g_{AB}$ depend only 
on the coordinates $r,x^3,\dots,x^n$ in
$\widetilde{\mathcal{O}}$. Notice first that since $\chi_u$ maps the
null hypersurface $\mathcal{N}$ into itself, geodesics into geodesics,
and preserves affine parametrisation, in particular, it is mapping all
the null geodesics starting at the points of $\mathcal{N}$ with
tangent $\mathfrak{L}^a$, which were used to set up our Gaussian null
coordinate system, into geodesics belonging to the same family.
Thereby, $\chi_u$ maps the $r=const, x^A=const$ coordinate lines, with
tangent $\mathfrak{K}^a$ into themselves.  Accordingly, $u$ is an
adapted Killing coordinate, i.e., all the smooth functions $f, h_{A}$
and $g_{AB}$ appearing in (\ref{le1}) have to be $u$-independent.

\medskip

Hereafter, we shall present our arguments only in domains where
Gaussian null coordinates can be defined as above.  It worth keeping
in mind, however, that an elementary spacetime neighbourhood
$\mathcal{O}$ can always be covered by sub-regions where Gaussian null
coordinates can be defined. By patching these type of coordinate
domains the ``local'' results derived in one of these coordinate
domains can always be seen to be valid in the associated elementary
spacetime region.

\section{The topology of the event horizon}\label{top}
\setcounter{equation}{0}

This section is to explore all those restrictions that follow from the
field equations on the possible topological properties of smooth
cross-sections of the event horizon of generic distorted stationary black
hole spacetimes.

Let us start by recalling that in case of asymptotically flat four
dimensional  stationary black hole spacetimes there is a fundamental
result, due to also Hawking \cite{hawk1}, asserting that, whenever the
dominant energy condition is satisfied smooth cross-sections of the
event horizon necessarily possesses the topology $S^2$. This result of
Hawking has been generalised  to the case of asymptotically (locally)
anti-de-Sitter spacetimes (see, e.g., \cite{G2,JV} and also
\cite{ga1,ga2} for higher dimensional generalisations). Recall that
under the assumption of asymptotic  flatness and the dominant
energy condition for matter fields (with zero cosmological constant)
it was also proved by Gannon \cite{gannon} that a smooth cross-section
of the event horizon of  a ``strongly future asymptotically
predictable'' black hole must be either a $2$-sphere or a torus.  Our
aim in this section is to carry out the analogous investigations
whenever no restriction concerning the asymptotic behaviour of the
black hole configurations is imposed. It will be shown that the
topology of the event horizon of the generic ``distorted'' black holes
is not as restricted as that of the aforementioned isolated
black hole configurations.

The following two lemmas are immediate consequences of the fact that
${\mathcal{N}}$ is a Killing horizon with respect to the Killing
vector field $\mathfrak{K}^a=(\partial /\partial u)^a$, which also
implies, in particular, that the null geodesic generators of
$\mathcal{N}$ are expansion and shear free. Notice that in this case
the space of the Killing orbits on $\mathcal{N}$ may also be endowed
with the natural Riemannian structure $(\mathcal{Z},-g_{AB})$.

\begin{lemma} \label{lemma1} 
Suppose that $(M,g_{ab})$ is a spacetime of class A.  Then, the
contraction $T_{ab}\mathfrak{K}^a\mathfrak{K}^b$ is identically zero
on ${\mathcal{N}}$. 
\end{lemma} 
  
\noindent\textbf{Proof}{\ } Since $\mathfrak{K}^a$ is null on
${\mathcal{N}}$ it follows from the Einstein's equations (\ref{ein})
that
\begin{equation} 
R_{ab}\mathfrak{K}^a\mathfrak{K}^b=8\pi
T_{ab}\mathfrak{K}^a\mathfrak{K}^b  \label{ein1} 
\end{equation} 
holds on ${\mathcal{N}}$. This, along with the fact that
$R_{ab}\mathfrak{K}^a\mathfrak{K}^b$ is identically zero on
${\mathcal{N}}$ (which can be verified by a direct calculation
carried out in the above introduced Gaussian null coordinates and by
making use of the $u$-independentness of the functions $f, h_{A}$ and
$g_{AB}$) justifies then the above claim.  \hfill \fbox{}

\bigskip

Since the Killing vector field $\mathfrak{K}^a$ is normal to
$\mathcal{N}$ and $\mathfrak{K}^a\mathfrak{K}_a=0$ on $\mathcal{N}$
there must exist a                              function\footnote{%
This function, playing the role of surface gravity, is denoted by
$\kappa_\circ$ not to be confused with the spin-coefficient $\kappa$
of the Newman-Penrose formalism \cite{newman:penrose} which will be
applied in our later discussions.}  $\kappa_{\circ}:\mathcal{N}
\rightarrow \mathbb{R}$ defined by the following equation
\begin{equation}\label{kappa}  
\frac12\nabla^a\left(\mathfrak{K}^e\mathfrak{K}_e\right)=
-\kappa_{\circ}\mathfrak{K}^a.    
\end{equation}
As an immediate consequence of (\ref{kappa}) we have that
$\kappa_{\circ}$ is constant along the null geodesic generators of
$\mathcal{N}$. The following lemma justifies that under the assumption
that the Einstein's equations are satisfied and the dominant energy
condition for matter fields holds the value of $\kappa_{\circ}$  does not
change from generator to generator.

\begin{lemma} \label{lemma2} 
Suppose that $(M,g_{ab})$ is a spacetime of class A.  Then, the
function $\kappa_{\circ}$ is constant throughout ${\mathcal{N}}$,
moreover, the  vector ${T^a}_{b}\mathfrak{K}^b$ points in the
direction of $\mathfrak{K}^a$ 
on ${\mathcal{N}}$.
\end{lemma} 
  
\noindent\textbf{Proof}{\  }  Let $(\widetilde{\mathcal{O}},g_{ab}
\vert_{\widetilde{\mathcal{O}}})$ be a neighbourhood of
$\widetilde{\mathcal{N}}\subset \mathcal{N}$, 
covered by Gaussian null coordinates
$\left(u,r,x^3,\dots,x^n\right)$.  Since  $\mathfrak{K}^a$
is normal to the coordinate basis field $\left(\partial /\partial
x^A\right)^a$  on $\widetilde{\mathcal{N}}$ it follows from the
Einstein's equations (\ref{ein}) that
\begin{equation} 
R_{ab}\mathfrak{K}^a\left(\frac{\partial }{\partial x^A}\right)^b= 8\pi
T_{ab}\mathfrak{K}^a\left( \frac{\partial }{\partial x^A}\right)^b
\label{ein2}
\end{equation} 
there. Moreover, since matter fields are assumed to satisfy the
dominant energy condition the vector field ${T^a}_b\mathfrak{K}^b$ has
to be future directed timelike or null on ${\mathcal{N}}$.  On the
other hand,  as we have seen above
$T_{ab}\mathfrak{K}^a\mathfrak{K}^b=0$ on ${\mathcal{N}}$ which
implies that ${T^a}_{b}\mathfrak{K}^b$ must point in the direction of
$\mathfrak{K}^a$.

It follows then that $T_{ab}\mathfrak{K}^a\left(\partial/\partial
x^A\right)^b$ has to vanish on $\widetilde{\mathcal{N}}$. Hence, by
making use of the $u$-independentness of the functions $f, h_{A}$ and
$g_{AB}$ we get, in the underlying Gaussian null coordinates, that
(\ref{ein2}) simplifies to
\begin{equation} 
\frac{\partial f}{\partial x^A}=0,  \label{ein2g} 
\end{equation} 
on $\widetilde{\mathcal{N}}$.  In virtue of (\ref{le1}) and
(\ref{kappa}) we have that $\kappa_{\circ}=-\frac12 f$, which along
with (\ref{ein2g}), implies then that $\kappa_{\circ}$ has to be
constant throughout $\widetilde{\mathcal{N}}$. Obviously, the constant
value of $\kappa_{\circ}$ has to be the same on overlapping Gaussian
coordinate domains which justifies, finally, the above claim. \hfill \fbox{}
   
\bigskip  

In consequence of (\ref{kappa}) we also have that
$\mathfrak{K}^a=(\partial/\partial   u)^a$  satisfies
\begin{equation}\label{sg}     
\mathfrak{K}^a\nabla_a\mathfrak{K}^b=\kappa_{\circ}\mathfrak{K}^b 
\end{equation} 
on  ${\mathcal{N}}$.  Thereby, the null geodesic generators of
${\mathcal{N}}$---which are complete with  respect to the parameter
$u$---are null geodesically complete  if $\kappa_{\circ }=0$ (in this
case the horizon is called to be degenerate), whereas, if
$\kappa_{\circ}$ happens to be nonzero the  generators  of
${\mathcal{N}}$ are geodesically complete only in one direction. In
the later case both the event horizon ${\mathcal{N}}$ and the
associated generic black hole spacetime are  called to be
non-degenerate. 

Motivated by the observation that the black hole region of a
stationary isolated  black hole is always bounded by a future event
horizon, hereafter, we shall assume that ${\mathcal{N}}$ is a future
event                                            horizon.\footnote{%
We would like to emphasise that analogous arguments---leading to the
same conclusions as derived below---apply whenever ${\mathcal{N}}$ is
assumed to be a past event horizon.} This implies, in particular, that
in the non-degenerate case ${\mathcal{N}}$ is generated by null
geodesics which are geodesically complete to the future but incomplete
to the past, moreover, we also have that $\kappa_{\circ}=-\frac12
f\vert_{\mathcal{N}}$ is a positive real number.

It is also important to note that Lemma\,\ref{lemma2}, in particular,
relation  (\ref{ein2g}) justifies that the ``zeroth law'' of black
hole thermodynamics generalise straightforwardly to the distorted
black hole spacetimes studied here.

\begin{proposition}\label{TOP}
Suppose that $(M,g_{ab})$ is a spacetime of class A.  Then, on the
smooth global cross-section ${\mathcal{Z}}$ of ${\mathcal{N}}$ the
relation
\begin{equation}\label{topp}
\int_\mathcal{Z}R^{^{(n-2)}}\epsilon_{_{\mathcal{Z}}} \leq
\int_\mathcal{Z}\left[2\tilde\Lambda +
 \kappa_{\circ} \cdot g^{CD}\partial_r g_{CD}+\frac{1}{2}h_Eh^E\right]
\epsilon_{_{\mathcal{Z}}} 
\end{equation}
is satisfied, where $R^{^{(n-2)}}$ and $\epsilon_{_{\mathcal{Z}}}$
denote the scalar curvature and the volume element associated with the
negative definite $n-2$-metric $g_{AB}$ on ${\mathcal{Z}}$.
\end{proposition} 
   
\noindent\textbf{Proof}{\ } We shall prove this statement by making
use of the ``$ur$'' component of the Einstein tensor in the applied
Gaussian null coordinate system  $\left(u,r,x^3,\dots,x^n\right)$. It
can be verified by direct calculations that on $\mathcal{N}$ the
relation
\begin{equation}\label{Gur}
G_{ur}=-\frac12
\left(\nabla^{^{(n-2)}}_Ah^A+R^{^{(n-2)}}+\frac{f}{2}\cdot g^{CD}
\partial_r 
g_{CD}-\frac12 h_Eh^E  
\right) 
\end{equation} 
holds, where $\nabla^{^{(n-2)}}_A$ denotes the covariant derivative
associated with the negative definite $n-2$-metric $g_{AB}$, moreover,
in all the applied raising of indices this metric has been used.

Recall, then, that $\mathfrak{K}^a$ is null and future directed on
$\mathcal{N}$. Hence, in virtue of the dominant energy condition along
with the conclusion of Lemma\,\ref{lemma2},  we have that
${T^a}_{b}\mathfrak{K}^b$ is future directed and parallel to
$\mathfrak{K}^a$, i.e., there exists a non-negative function
$\varphi:\mathcal{N}\rightarrow \mathbb{R}$ so that
\begin{equation}
{T^a}_{b}\mathfrak{K}^b=\varphi \mathfrak{K}^a.
\end{equation} 
By contracting this relation with $\mathfrak{L}_a$, and taking into
account that 
$g_{ur}=\mathfrak{L}^a\mathfrak{K}_a=1$, we get   
\begin{equation}
T_{ur}={T}_{ab}\mathfrak{K}^a\mathfrak{L}^b=\varphi\geq 0.
\end{equation} 
This later relation, along with (\ref{Gur}) and the Einstein's equations
(\ref{ein}), implies then that  
\begin{equation}
G_{ur}+\tilde\Lambda g_{ur}=-\frac12
\left(\nabla^{^{(n-2)}}_Ah^A+R^{^{(n-2)}}+\frac{f}{2}\cdot g^{CD}\partial_r
g_{CD}-\frac12 h_Eh^E \right)+  \tilde\Lambda \geq 0.
\end{equation}
Since $\mathcal{Z}$ has been assumed to be an orientable compact
manifold with no boundary relation (\ref{topp}) immediately
follows then.  \hfill \fbox{} \bigskip 

\medskip

There are important consequences of the above result restricting the
topological character of the cross-sections of the event horizon of
``distorted'' black holes. To see this we need to investigate first
the contribution of the integral  
$\int_\mathcal{Z}
\kappa_{\circ}\cdot g^{CD}\partial_r g_{CD} \epsilon_{_{\mathcal{Z}}}$
to the {\it r.h.s.}\,of (\ref{topp}).
In doing so notice first that the term
$g^{CD}\partial_r g_{CD}$ can be related to the null expansion
$\theta_{(\mathfrak{L})}=\nabla^a\mathfrak{L}_a$ of the null geodesic
congruence with tangent field $\mathfrak{L}^a$ as
\begin{equation}
\theta_{(\mathfrak{L})}=g^{CD}\partial_r g_{CD}.
\end{equation}
Second, it follows from (\ref{kappa}) that if the event horizon
$\mathcal{N}$ is non-degenerate the norm of the Killing field change
sign on $\mathcal{N}$. Thereby, in the non-degenerate case,
$\mathfrak{K}^a$ has to be timelike on the side of $\mathcal{N}$, (at
least) in a sufficiently small open neighbourhood of $\mathcal{N}$,
which---in consequence of the fact that $\mathcal{N}$ was chosen to be
a future event horizon---has to belong to the chronological past
$I^-[\mathcal{N}]$ of $\mathcal{N}$ in
$\mathcal{O}$. We shall refer to the corresponding part of
$\mathcal{O}$ as the {\it domain of outer communication} with respect
to $\mathcal{N}$, and the associated region will be denoted by
$\mathcal{D}_\mathcal{N}$, if 
\begin{equation}\label{av0}
\int_\mathcal{Z}
\theta_{(\mathfrak{L})}\epsilon_{_{\mathcal{Z}}}\leq 0\,.
\end{equation} 

Notice that this definition of the domain of outer communication is
compatible with the following intuitive picture. Consider the null
hypersurface $\mathcal{N}^T$, intersecting $\mathcal{N}$ transversely
at $\mathcal{Z}$, which is generated by null geodesics starting at the
points of the cross-section $\mathcal{Z}$ with tangent
$\mathfrak{L}^a$. Since $\mathcal{N}^T$ is smooth in $\mathcal{O}$ it
can be smoothly foliated there by $(n-2)$-dimensional surfaces,
$\mathcal{Z}_r$, defined as the $r=const$ cross-sections of
$\mathcal{N}^T$. Then, if the ``area'' $\mathcal{A}(\mathcal{Z}_r)=
\int_{\mathcal{Z}_r}\epsilon_{_{\mathcal{Z}_r}}$ of these
cross-sections is non-decreasing towards the domain where
$\mathfrak{K}^a$ is timelike we  consider the associated domain as
being ``outer'' with respect to $\mathcal{N}$. To see that, in fact,
this simple geometric idea was applied in the above definition recall
that
\begin{equation}\label{av00}
\frac{{\rm d} \mathcal{A}(\mathcal{Z}_r)}{{\rm
    d}r}\vert_{{\mathcal{Z}}}= \int_\mathcal{Z}
    \pounds_\mathfrak{L}\left(\epsilon_{_{\mathcal{Z}}}\right)=
    \int_\mathcal{Z}
    \theta_{(\mathfrak{L})}\epsilon_{_{\mathcal{Z}}}\,,
\end{equation} 
moreover, that $\mathfrak{L}^a$ is future directed on $\mathcal{N}$.  

Hereafter we shall assume that $\mathcal{N}$ is either degenerate or
it is non-degenerate but lies on the boundary of  the domain of outer
communication, $\mathcal{D}_\mathcal{N}$.  In the latter case, the
cross-section $\mathcal{Z}$ may also be called to be ``{\it convex on
the average}'' from the direction of $\mathcal{D}_\mathcal{N}$. Notice
that even in this case $\mathcal{Z}$ need not to be everywhere convex,
i.e., there may be subsets of $\mathcal{Z}$ so that on these subsets
the null geodesics intersecting $\mathcal{Z}$ transversely are locally
converging towards the direction of $\mathcal{D}_\mathcal{N}$.
Nevertheless, the most important consequence of all above is that
whenever $\mathcal{N}$ is  either degenerate, with $\kappa_{\circ}=0$, or
it is non-degenerate, with $\kappa_{\circ}>0$, but $\mathcal{N}$ lies  on
the boundary of  the domain of outer communication the integral
$\int_\mathcal{Z} \kappa_{\circ}\cdot g^{CD}\partial_r g_{CD}
\epsilon_{_{\mathcal{Z}}}$ on the {\it r.h.s.}\,of (\ref{topp}) has to
be smaller than or equal to zero.

Taking, then, into account that $g_{AB}$ is negative definite and
summarising  what we have justified above the following result---which
is in accordance with the findings of Galloway and Schoen
\cite{ga1,ga2} in case of stationary black hole spacetimes with
everywhere convex horizons---can be seen to hold which also provides
generalisations of some of the results of \cite{HSN} for the higher
dimensional case.
\begin{corollary}\label{tipp1} 
Suppose that $(M,g_{ab})$ is a spacetime of class A. Assume,
furthermore, that $\mathcal{N}$ is either degenerate or it is
non-degenerate but lies on the boundary of  the domain of outer
communication $\mathcal{D}_\mathcal{N}$. Then, for any smooth global
cross-section $\mathcal{Z}$ of $\mathcal{N}$ we have that
\begin{equation}\label{tipp1e}
\int_\mathcal{Z}R^{^{(n-2)}}\epsilon_{_{\mathcal{Z}}}  \leq
2\tilde\Lambda\cdot \mathcal{A}(\mathcal{Z})\,.
\end{equation}
\end{corollary} 

In the particular case of a $4$-dimensional spacetime the relation 
$\mathcal{K}_\mathcal{G}=-\frac12 R^{^{(2)}}$ of the Gaussian and scalar
curvatures of a $2$-dimensional global cross-section $\mathcal{Z}$,
along with the Gauss-Bonnet theorem, implies that
\begin{equation}\label{tipp2e} 
2\pi\chi_{_\mathcal{Z}}=4\pi(1-g_{_\mathcal{Z}})=
\int_\mathcal{Z}\mathcal{K}_\mathcal{G} 
\epsilon_{_{\mathcal{Z}}}=
-\frac12\int_\mathcal{Z}R^{^{(2)}} 
\epsilon_{_{\mathcal{Z}}}\,,
\end{equation}
where $\chi_{_\mathcal{Z}}$ and $g_{_\mathcal{Z}}$ denote the Euler
characteristic and the ``genus'' of 
$\mathcal{Z}$, respectively. Thereby, as an
immediate consequence of Cor.\,\ref{tipp1} we also have.
\begin{corollary}\label{tipp2} 
Suppose that $(M,g_{ab})$ is a $4$-dimensional spacetime of
class A.  Assume, furthermore, that $\mathcal{N}$ is  either
degenerate or it is non-degenerate but lies on the boundary of  the
domain of outer communication $\mathcal{D}_\mathcal{N}$.  Then,
whenever the cosmological constant  (in our signature) is
non-positive, $\tilde\Lambda\leq 0$, any smooth global cross-section
$\mathcal{Z}$ of $\mathcal{N}$ must possess the topology of either a
sphere or a torus, while $\mathcal{Z}$ may be a compact orientable
$2$-surface of genus $g_{_\mathcal{Z}}> 1$ 
whenever $\tilde\Lambda> 0$.
\end{corollary} 

We would like to emphasise that in case of degenerate horizons (with
$\kappa_{\circ}=0$) the conclusions of the above corollaries remain
valid regardless whether $\mathcal{N}$ lies on the boundary of a
domain of outer communication or not.

Let us finally consider a spacetime of class A with non-positive
cosmological constant, $\tilde\Lambda\leq 0$, and assume that it
possesses a horizon with a cross-section $\mathcal{Z}$ so that
$\int_\mathcal{Z}R^{^{(n-2)}} \epsilon_{_{\mathcal{Z}}}=0$. Notice
that then besides the conclusion of Cor.\,\ref{tipp1} it also follows
immediately from (\ref{topp}) and  (\ref{tipp1e}) that $\tilde\Lambda$
has to be zero, moreover, either  $\kappa_{\circ}$ or
$\int_\mathcal{Z}\theta_{(\mathfrak{L})}$, as well as, $h_E$ have to
vanish identically on $\mathcal{N}$.  This, in the particular case of
a $4$-dimensional spacetime, implies that if the cosmological
constant is strictly negative, then any smooth global cross-section
$\mathcal{Z}$ of $\mathcal{N}$ must possess the topology of a
$2$-sphere.

\section{Distorted black hole spacetimes of class B}\label{pre2}
\setcounter{equation}{0}

As we have seen above the null geodesic generators of the future event
horizon of a (non-degenerate) distorted black hole spacetime are
geodesically  incomplete to the past. Appealing then to the results of
\cite{rw1,rw2} (see also \cite{frw,r1}) it can be  shown that to any
(non-degenerate) black hole spacetime $(M,g_{ab})$ of class A the open
neighbourhood ${\mathcal{O}}$ of the (future) event horizon,
$\mathcal{N}$, can always be chosen to be sufficiently small  so that
the subspacetime $(\mathcal{O},g_{ab}|_{\mathcal{O}})$,  considered
now as a spacetime on its own right, can be extended. In particular,
the existence of a smooth extension $(\mathcal{O}^*,g^*_{ab})$ of
$(\mathcal{O},g_{ab}|_{\mathcal{O}})$ can be shown so that
$(\mathcal{O}^*,g^*_{ab})$ possesses a bifurcate null surface,
${\mathcal{H}}^*$---i.e., ${\mathcal{H}}^*$ is the union of two null
hypersurfaces, $\mathcal{H}_1$ and $\mathcal{H}_2$, which intersect on
an  $n-2$-dimensional spacelike surface, $\mathcal{Z}$---such that
${\mathcal{N}}$ corresponds to the portion of $\mathcal{H}_1$ that
lies to the causal future of $\mathcal{Z}$.

Furthermore, in the particular case of a (non-degenerate) black hole
spacetime $(M,g_{ab})$ of class B, i.e., in case of a $4$-dimensional
electrovac black hole spacetime it can also be guaranteed that the
extended spacetime $(\mathcal{O}^*,g^*_{ab})$ possesses a ``wedge
reflection'' symmetry, moreover, the vector field $\mathfrak{K}^a$ and
the electromagnetic field $F_{\,ab}$ extend from $\mathcal{O}$ to
fields ${\mathfrak{K}^*}^a$ and $F_{\,ab}^*$ on $\mathcal{O}^*$ so
that the Lie derivatives $\mathcal{L}_{\mathfrak{K}^*}g^*_{ab}$ and
$\mathcal{L}_{\mathfrak{K}^*} F_{\,ab}^*$ vanish identically on
$\mathcal{O}^*$ \cite{rw2} (see also \cite{r1,rkill,rkill2}).

\medskip 

Due to the extension process it can easily be 
seen that the null generators of both of the null hypersurfaces,
$\mathcal{H}_1$ and $\mathcal{H}_2$, comprising the bifurcate type
event horizon ${\mathcal{H}}^*$, are geodesically complete, i.e., they
extend arbitrary large values of their affine parameters. Starting
now, for instance, by the null geodesically complete hypersurface
$\mathcal{H}_1$ one may repeat the construction yielding an
elementary spacetime neighbourhood and associated Gaussian null
coordinate systems. It is straightforward to verify---by recalling the
details of the extension process (see \cite{rw1,rw2})---that the
entire of $\mathcal{O}^*$  immediately gives rise to an elementary
spacetime neighbourhood of $\mathcal{H}_1$.

\medskip

In the rest of this paper we have restrict our considerations to
``distorted'' electrovac black hole spacetimes of class B.  Since
hereafter we shall frequently refer to results of
\cite{newman:penrose} and \cite{friedrich} both of which works refer
to the Gaussian null coordinates $\left(u,r,x^3,x^4\right)$ in a
slightly different context than we have done in the previous part of
this paper before proceeding we shall make the following simple
changes in our notations.  Hereafter the formerly defined Gaussian
null coordinates will be ``hatted''. In particular, the Killing
parameter associated with the horizon Killing vector field
$\mathfrak{K}^a$ will be denoted by $\hat u$. Similarly, the affine
parameter along the null geodesics starting at the points of
$\mathcal{N}$ with tangent $\mathfrak{L}^a$ and  synchronised so that
it vanishes on $\mathcal{N}$ will be denoted as $\hat r$. Adopting
this notation a synchronised affine parametrisation of the null
geodesic generators of $\mathcal{N}=J^+[\mathcal{Z}] \cap
\mathcal{H}_1$ can then be given as
\begin{equation}
u=e^{\kappa_\circ \hat u}.
\end{equation}      
The associated parallelly propagated tangent vector field
$k^a=\left(\partial/\partial u\right)^a$ on $\mathcal{N}$ can be
related to $\mathfrak{K}{}^a$ as  
\begin{equation}\label{kur}
k^a=\frac{1}{\kappa_\circ}e^{-\kappa_\circ
  \hat u}\mathfrak{K}^a=\frac{1}{\kappa_\circ u}\mathfrak{K}^a.
\end{equation} 
According to this choice the affine parameter $u$ takes positive
values, $u>0$ on $\mathcal{N}$, while the associated  Killing
parameter $\hat u$ runs from $-\infty$ to $\infty$. Nevertheless,  $u$
can be extended immediately onto $\mathcal{H}_1$ so that it will be an
affine parameter everywhere along the null geodesic generators of
$\mathcal{H}_1$ and also which has been synchronised so that $u=0$
corresponds to $\mathcal{Z}$. 

By repeating the basic points of the construction applied in
Section\,\ref{pre},  based on  the  $1$-parameter family of smooth
cross-sections $\mathcal{Z}_u=\{p\in\mathcal{H}_1\vert u(p)=u\in
\mathbb{R} \}$ of $\mathcal{H}_1$ the base manifold $\mathcal{O}^*$
can be seen to be an {\it elementary spacetime region}. Moreover, {\it
Gaussian null coordinates}, $\left(u,r,x^3,x^4\right)$, can be
introduced on suitable subsets, ${\widetilde{\mathcal{O}}}$, of
$\mathcal{O}^*$. The most significant differences are the
following. First,  the components of the metric, the smooth functions
$f, h_A$ and $g_{AB}$, appearing in the line element (\ref{le1}) will
not be independent of the $u$-coordinate as for the coordinate basis
field $k^a=\left(\partial/\partial u\right)^a$ need not to be a
Killing vector field. Second, while $\mathcal{O}^*$ can be covered by
the Gaussian null coordinate patches, $\{{\widetilde{\mathcal{O}}}\}$,
the formerly defined Gaussian null 
coordinates $\left(\hat u,\hat r,\hat x^3,\hat x^4\right)$, based on
the use of the horizon Killing vector field
$\mathfrak{K}^a=(\partial/\partial {\hat u})^a$, can be defined only
on the part, ${\widetilde{\mathcal{O}}_+}$, of each
${\widetilde{\mathcal{O}}}$ with $u>0$.

It can be checked then that   the  coordinates ``\,$r$\,'' and
``\,$\hat r$\,'', which are affine parameters along the null geodesics
with tangents $l^a$ and $\mathfrak{L}^a$ in
${\widetilde{\mathcal{O}}_+}$, respectively, can---in virtue of the
relations $l^a=g^{ab}\nabla_b u$ and $\mathfrak{L}^a=g^{ab}\nabla_b
\hat u$---be related as
\begin{equation}
\hat r=\kappa_\circ u r.
\end{equation} 

\subsection{The Newman-Penrose formalism}\label{npfr}

In deriving the main results of the rest of this paper we are going to
apply a combination of the Newman-Penrose formalism
\cite{newman:penrose} and the null characteristic initial value
formulation of Einstein's theory of gravity, as it was worked out in
details by Friedrich \cite{friedrich} (see also
\cite{friedrich1,friedrich2} for related investigations). Thereby, it
seems to be useful to recall the relation between the geometrical
setting associated with the above introduced Gaussian null coordinates
and the fundamental layout of the Newman-Penrose formalism
\cite{newman:penrose} and that of \cite{friedrich} which is done in
this subsection.
  
\medskip

The contravariant form of the spacetime metric, given in the form of
(\ref{le1}) in a Gaussian null coordinate  system $(u,r,x^3,x^4)$
covering the part $\widetilde{\mathcal{O}}$  of an elementary
spacetime region ${\mathcal{O}}$, can be given as
\begin{equation} 
g^{\alpha\beta}=\left( 
\begin{array}{ccc} 
0 & 1 & 0 \\ 
1 & g^{rr} & g^{rB} \\ 
0 & g^{Ar} & g^{AB} 
\end{array} 
\right) .  \label{m2} 
\end{equation} 
Choosing now, as it was done in \cite{newman:penrose}, real-valued
functions $U $, $X ^A$ and complex-valued 
functions $\omega ,$ $\xi ^A$ on $\widetilde{\mathcal{O}}$ such that 
\begin{equation} 
g^{rr}=2(U -\omega \bar\omega),\ \ g^{rA}=X ^A- (\bar\omega\xi 
^A+\omega \bar\xi^A),\ \ g^{AB}=-(\xi ^A \bar\xi^B+\bar\xi^A\xi ^B), 
\label{m3} 
\end{equation} 
and setting 
\begin{equation} 
l^\mu =\delta ^\mu {}_r,\ \ n^\mu =\delta ^\mu {}_u+U \delta ^\mu 
{}_r+X ^A\delta ^\mu {}_A, \ \ m^\mu =\omega \delta ^\mu {}_r+\xi ^A\delta 
^\mu {}_A,  \label{tet} 
\end{equation} 
we obtain a complex null tetrad $\{l^a,n^a,m^a,\overline{m}^a\}$ in
$\widetilde{\mathcal{O}}$. We require   that $U $, $X ^A$, and $\omega
$ vanish on $\widetilde{\mathcal{H}}_1$  which guaranties that $n^a$
is tangent to the generators of $\widetilde{ \mathcal{H}}_1$,
$n^a=k^a$ there, moreover, $m^a$ and $\overline{m}^a$ are everywhere
tangent to the cross-sections $\widetilde{\mathcal{Z}}_u$ of
$\widetilde{ \mathcal{H}}_1$. In the following we shall consider the
derivatives of  functions in the direction of the frame vectors
defined above and denote the  corresponding operators in
$\widetilde{\mathcal{O}}$ by
\begin{equation} 
\mathrm{D}=\partial /\partial r,\ \ \Delta =\partial /\partial u+ U \cdot
\partial /\partial r+X^A \cdot\partial /\partial x^A,\ \  \delta 
=\omega \cdot\partial /\partial r+\xi ^A\cdot\partial /\partial x^A. 
\label{dop} 
\end{equation} 
To simplify the Newman-Penrose equation a part of the remaining gauge
freedom can be fixed, as it was already done in \cite{newman:penrose},
by assuming that the tetrad $\{l^a,n^a,m^a,\overline{m}^a\}$ is
parallelly propagated along the null geodesics with tangent
$l^a=\left( \partial /\partial r\right) ^a$ in
$\widetilde{\mathcal{O}}$. These assumptions guarantee that for the
spin coefficients, corresponding to this specific choice of complex
null tetrad,  $\kappa =\pi =\varepsilon =0$, $\rho = \overline{\rho
}$, $\tau =\overline{\alpha }+\beta$ hold everywhere in
$\widetilde{\mathcal{O}}$. Moreover, since we have chosen $n^a$ so
that $n^e\nabla_e n^a=0$ along the generators of
$\widetilde{\mathcal{H}}_1$ the spin coefficient $\nu$, by its
definition, is guaranteed  to vanish on
$\widetilde{\mathcal{H}}_1$. Finally, since $u$ is an affine parameter
along the generators of $\widetilde{\mathcal{H}}_1$, e.g., in virtue
of (4.14) of \cite{r1}, we also have that $\gamma+\overline\gamma=0$
thereon. In this case we can also apply a rotation of the form $m^a
\rightarrow e^{i\phi}m^a$, where  $\phi: \widetilde{\mathcal{H}}_1
\rightarrow\mathbb{R}$ is a suitably chosen real function, so that the
spin coefficient $\gamma$ will, in turn, vanish everywhere on
$\widetilde{\mathcal{H}}_1$.

\subsection{The null characteristic formulation}\label{nchfr}

We would like to emphasise that the gauge choices we have made so far
are exactly the same as those were used in \cite{friedrich}, hence,
all of the results of Friedrich's formalism can be applied. Here we
start by the investigation of the pure vacuum case with vanishing
cosmological constant. Later, in
Section\,\ref{EM}, it will be shown how the techniques applied below
extend to spacetimes with non-zero cosmological constant and with a
source free electromagnetic field.

\medskip

Recall first that the pertinent Newman-Penrose equations\footnote{%
To avoid the steady citation of this fundamental work of Newman and
Penrose \cite{newman:penrose}  throughout this paper the equations
referred to as $(NP.6.$`a combination of a number $\&$ a lowercase
letter'$)$ are always meant to be the original equations listed as
$($6.`a combination of a number $\&$ a lowercase letter'$)$ in
\cite{newman:penrose}. }, (NP.6.10a)-(NP.6.10h), (NP.6.11a)-(NP.6.11r)
and (NP.6.12a)-(NP.6.12h), taking them as first order partial
differential equations, with  respect to Gaussian null coordinates,
$(u,r,x^3,x^4)$ in $\widetilde{\mathcal{O}}$, for
the vector valued variable
\begin{equation}\label{V} 
\mathbb{V}=(\xi^A,\omega,X^A,U;\rho,\sigma,\tau,\alpha,\beta,
\gamma,\lambda,
\mu,\nu;\Psi_0,\Psi_1, \Psi_2,\Psi_3,\Psi_4)
\end{equation}
are overdetermined simply because there are more equations than
unknowns. Nevertheless, as it was proved by Friedrich
\cite{friedrich}, by taking aside some of the Newman-Penrose equations
and taking linear combinations some of them the following ``reduced
set of vacuum field equations''\footnote{%
To distinguish these equations from others applied in this paper,
we shall label the $n^{th}$ reduced equation as ``$($FR.$n)$''.  }
\renewcommand{\theequation}{FR.\arabic{equation}}
\setcounter{equation}{0}
\begin{eqnarray} 
&&\hskip-.0cm \mathrm{D}\,\xi ^A = {\rho }\,\xi ^A+{\sigma
}\,\overline\xi^A \\  &&\hskip-.0cm \mathrm{D}\,\omega = {\rho\,
\omega }+\sigma \,\overline{\omega }-\tau \\  &&\hskip-.0cm
\mathrm{D}X^A = {\tau }\,\overline{\xi}^A+{\overline{\tau}\,\xi }^A \\
&&\hskip-.0cm  \mathrm{D}\,U = {\tau }\,\overline{\omega
}+{\overline{\tau}\,\omega -(\gamma +}  \overline{\gamma }) \\ &&
\mathrm{D}\,{\rho } = {\rho }^2+{\sigma \,\overline{\sigma}}\\  &&
\mathrm{D}\,{\sigma } = 2{\rho \,\sigma }+{{\Psi }_0}  \\  &&
\mathrm{D}\,{\tau } = {\tau \,\rho }+{\overline{\tau}\,\sigma }+{{\Psi
}_1} \\  &&
\mathrm{D}\,{\alpha} = {\rho \,\alpha }+{\beta
\,\overline{\sigma}} \\  && \mathrm{D}\,\beta = {\alpha \,\sigma
}+{\rho \,\beta }+{{\Psi }_1} \\  && \mathrm{D}\,{\gamma} = {\tau
\,\alpha }+{\overline{\tau}\,\beta }+{{\Psi }_2} \\  &&
\mathrm{D}\,{\lambda} = {\rho\, \lambda }+{\overline{\sigma}\,\mu } \\
&& \mathrm{D}\,{\mu } = {\rho \,\mu }+{\sigma\,\lambda }+{{\Psi }_2}\\
&& \mathrm{D}\,{\nu} =
{\overline{\tau}\,\mu}+{\tau\,\lambda}+{{\Psi}_3} \\ &&{\hskip-2.0cm
\Delta\,\Psi_0-\delta\,\Psi _1 = (4\,\gamma -\mu )\,\Psi
_0-2\,(2\,\tau+\beta )\,\Psi_1+3\,\sigma\,\Psi_2}  \\ &&{\hskip-2.0cm
\Delta\,\Psi _1+\mathrm{D}\,\Psi_1-\delta\,\Psi _2-\overline{\delta
}\,\Psi _0 =   (\nu-4\,\alpha)\,\Psi_0-2\,(\mu -\gamma
-2\,\rho)\,\Psi_1-3\,\tau\,\Psi_2+2\,\sigma\,\Psi_3} \\ &&{\hskip-2.0cm
\Delta\,\Psi_2 + \mathrm{D}\,\Psi_2-{\delta }\,\Psi
_3-\overline{\delta }\,\Psi _1 = -\lambda  \,\Psi_0-2\,(\alpha-\nu)\,
\Psi _1+3\,(\rho-\mu)\,\Psi _2 - 2\,(\tau-\beta)\, \Psi _3
+\sigma\,\Psi_4}\\ &&{\hskip-2.0cm  \Delta\,\Psi_3 +
\mathrm{D}\,\Psi_3-{\delta }\,\Psi _4-\overline{\delta }\,\Psi _2
=-2\,\lambda \,\Psi_1 + 3\,\nu\,\Psi_2 + 2\,(\rho -\gamma
-2\,\mu)\,\Psi_3 + (4\,\beta-\tau)\,\Psi_4} \\ &&{\hskip-2.0cm
\mathrm{D}\,\Psi_4-\overline{\delta }\,\Psi _3 = -3\,\lambda \,\Psi_2
+ 2\,\alpha\,\Psi_3 + \rho\,\Psi_4}
\end{eqnarray}
can be derived. These equations, besides
constituting a determined system for the vector variable,
$\mathbb{V}$, are as good as the complete set of the
Newman-Penrose equations. More precisely, what was proved by Friedrich
(see Theorem 1.\,of \cite{friedrich}) can be rephrased as.
\begin{theorem}\label{HF} 
Denote by $\mathbb{V}_0$ an initial data set, satisfying the ``inner''
Newman-Penrose equations on the initial data surface comprised by the
pair of intersecting null hypersurfaces $\widetilde{\mathcal{H}}_1$
and $\widetilde{\mathcal{H}}_2$. If $\mathbb{V}$ is the solution on
the domain of dependence $D[\widetilde{\mathcal{H}}_1\cup
\widetilde{\mathcal{H}}_2]$ to the reduced vacuum field equations with
$\mathbb{V}\vert_{{\widetilde{\mathcal{H}}_1\cup
\widetilde{\mathcal{H}}_2}}=\mathbb{V}_0$, then $\mathbb{V}$ is also a
solution to the full set of the Newman-Penrose equations. Moreover,
the metric, the connection and the curvature tensor determined by
$\mathbb{V}$ are so that the connection will be metric and torsion
free, as well as, the curvature tensor which can be built from the
Weyl spinor components is the curvature tensor associated with this
torsion free connection.
\end{theorem} 

\renewcommand{\theequation}{5.\arabic{equation}}
\setcounter{equation}{8}

We would like to emphasise that the condition requiring the initial
data $\mathbb{V}_0$ to satisfy the ``inner'' Newman-Penrose equations
on $\widetilde{\mathcal{H}}_1 \cup \widetilde{\mathcal{H}}_2$ is not
as restrictive as it seems to be. Indeed,  if we are given a pair of
smooth null hypersurfaces $\widetilde{\mathcal{H}}_1$ and
$\widetilde{\mathcal{H}}_2$ intersecting on a $2$-dimensional
spacelike surface $\widetilde{\mathcal{Z}}$, some of the
Newman-Penrose equations will be ``interior equations'' on
$\widetilde{\mathcal{Z}}$, $\widetilde{\mathcal{H}}_1$ and
$\widetilde{\mathcal{H}}_2$, respectively. Therefore, as it was shown
by Friedrich, see the argument related to Lemma 1.\,in
\cite{friedrich}, we may start with a ``reduced initial data set'',
$\mathbb{V}_0^{red}$, which consists of the specification of the Weyl
spinor components $\Psi_4$ on $\widetilde{\mathcal{H}}_1$ and $\Psi_0$
on $\widetilde{\mathcal{H}}_2$, moreover, it includes 
the specification of the spin-coefficients
$\rho,\sigma,\tau,\mu,\lambda$, along with a vector field $\xi^A$ such
that $g^{AB}=-(\xi^A\overline{\xi}^B+\overline{\xi}^A\xi^B)$ is a
negative definite metric, on $\widetilde{\mathcal{Z}}$. It is argued
then that the ``inner equations on $\widetilde{\mathcal{Z}}$'' can be
solved algebraically for the rest of the variables listed in
$\mathbb{V}$. Moreover, once the components of $\mathbb{V}$ are known
on $\widetilde{\mathcal{Z}}$ the desired initial data $\mathbb{V}_0$
can be determined on $\widetilde{\mathcal{H}}_1$ and
$\widetilde{\mathcal{H}}_2$ by integrating a sequence of ordinary
differential equations---these are the corresponding inner
equations---along the null geodesic generators a
$\widetilde{\mathcal{H}}_1$ and $\widetilde{\mathcal{H}}_2$,
respectively. Notice that this $\mathbb{V}_0$, by construction,
satisfies all the inner equations as it was assumed in
Theorem\,\ref{HF} above. The way $\mathbb{V}_0$ is determined---in the
particular case of a bifurcate Killing horizon, 
$\widetilde{\mathcal{H}}^*=\widetilde{\mathcal{H}}_1 \cup
\widetilde{\mathcal{H}}_2$---will be illustrated in
Subsection\,\ref{app}.  What will be important for us in our later
investigations is the following result of Friedrich \cite{friedrich}.
\begin{lemma}\label{lHF2} 
Assume that $\mathbb{V}$ is a solution to the Newman-Penrose
equations. Denote by $\mathbb{V}_0$ the restriction of $\mathbb{V}$
onto $\widetilde{\mathcal{H}}_1 \cup \widetilde{\mathcal{H}}_2$,
moreover, by $\mathbb{V}_0^{red}$ the corresponding reduced initial
data as specified above. Then $\mathbb{V}_0^{red}$, along with the
Newman-Penrose equations, determines uniquely the initial data set
$\mathbb{V}_0$ on $\widetilde{\mathcal{H}}_1 \cup
\widetilde{\mathcal{H}}_2$.
\end{lemma}

In addition to the fact that the above reduced vacuum field equations,
(FR.1)-(FR.18), comprise a determined system, it can also be verified
that, when written out in Gaussian null coordinates $(u,r,x^3,x^4)$ in
$\widetilde{\mathcal{O}}$, they possess the form
\begin{equation}\label{eqV}
\mathbb{A}^\mu \cdot \partial_\mu \mathbb{V} + \mathbb{B}=0,
\end{equation}
where the matrices $\mathbb{A}^\mu$ and $\mathbb{B}$ smoothly depend
on $\mathbb{V}$, along with its complex conjugate
$\overline{\mathbb{V}}$. Moreover, it can also be seen that the
matrices $\mathbb{A}^\mu$ are Hermitian, i.e.,
$\overline{\mathbb{A}}{}^\mu{}^T={\mathbb{A}}^\mu$ and the combination
$\mathbb{A}^\mu(n_\mu+l_\mu)$ is positive definite at least in a
sufficiently small neighbourhood of $\widetilde{\mathcal{H}}_1$.

The validity of the latter assertions can be justified by direct
inspection of equations (FR.1)-(FR.18). Notice first that the
coefficient matrices of 
the derivative  operators $\rm{D},\Delta,\delta,\overline\delta$ have
the form of $18\times18$ matrices given as
\begin{equation}\label{M1}
\mathbb{A}^{\rm{D}}=\left( {\begin{array}{c|c}
\mathbf{1} & \mathbf{0}\\ \hline 
\mathbf{0} &  \begin{array}{rrrrr}
 0 & 0 & 0 & 0 & 0 \\
 0 & 1 & 0 & 0 & 0 \\
 0 & 0 & 1 & 0 & 0 \\
 0 & 0 & 0 & 1 & 0 \\
 0 & 0 & 0 & 0 & 1
\end{array} 
\end{array}}
 \right) 
\ \ \ \ \ 
\mathbb{A}^{\Delta}=\left( {\begin{array}{c|c}
\mathbf{0} & \mathbf{0}\\ \hline 
\mathbf{0} &  \begin{array}{rrrrr}
 1 & 0 & 0 & 0 & 0 \\
 0 & 1 & 0 & 0 & 0 \\
 0 & 0 & 1 & 0 & 0 \\
 0 & 0 & 0 & 1 & 0 \\
 0 & 0 & 0 & 0 & 0
\end{array} 
\end{array}}
 \right) 
\end{equation}
\begin{equation}\label{M2}
\mathbb{A}^{\delta}=\left( {\begin{array}{c|c}
\mathbf{1} & \mathbf{0}\\ \hline 
\mathbf{0} &  \begin{array}{rrrrr}
 0 & -1 & 0 & 0 & 0 \\
 0 & 0 & -1 & 0 & 0 \\
 0 & 0 & 0 & -1 & 0 \\
 0 & 0 & 0 & 0 & -1 \\
 0 & 0 & 0 & 0 & 0
\end{array} 
\end{array}}
 \right)
\ \ \ \ \ 
\mathbb{A}^{\overline\delta}=\left( {\begin{array}{c|c}
\mathbf{0} & \mathbf{0}\\ \hline 
\mathbf{0} &  \begin{array}{rrrrr}
 0 & 0 & 0 & 0 & 0 \\
 -1 & 0 & 0 & 0 & 0 \\
 0 & -1 & 0 & 0 & 0 \\
 0 & 0 & -1 & 0 & 0 \\
 0 & 0 & 0 & -1 & 0
\end{array} 
\end{array}}
 \right)\,, 
\end{equation}
where $\mathbf{1}$ stands for the $13\times13$ identity matrix while
$\mathbf{0}$ always denotes suitable type of matrices with identically
zero elements.    
Taking then into account the decomposition 
\begin{equation}
\mathbb{A}^\mu \cdot \partial_\mu=\mathbb{A}^{\rm{D}} \cdot \rm{D} +
\mathbb{A}^{\Delta} \cdot \Delta + \mathbb{A}^{\delta} \cdot \delta +
\mathbb{A}^{\overline\delta} \cdot \overline\delta\,, 
\end{equation}
along with the relations (\ref{dop}), expressing the derivative 
operators $\rm{D},\Delta,\delta,\overline\delta$ in terms of the
partial derivatives with respect to the Gaussian null coordinates
$(u,r,x^3,x^4)$ in $\widetilde{\mathcal{O}}$, we get that
\begin{eqnarray}
&&\mathbb{A}^u=\mathbb{A}^{\Delta}\\
&&\mathbb{A}^r=\mathbb{A}^{\rm{D}}+U\cdot\mathbb{A}^{\Delta}
  +\omega\cdot\mathbb{A}^{\delta}
  +\overline\omega\cdot\mathbb{A}^{\overline\delta}\\  
&&\mathbb{A}^A=X^A\cdot\mathbb{A}^{\Delta}
  +\xi^A\cdot\mathbb{A}^{\delta}
  +\overline\xi^A\cdot\mathbb{A}^{\overline\delta}\,.  
\end{eqnarray} 
It is straightforward to see then, in virtue of the explicit forms of
the matrices $\mathbb{A}^{\rm{D}}$, $\mathbb{A}^{\Delta}$,
$\mathbb{A}^{\delta}$ and $\mathbb{A}^{\overline\delta}$ given by
(\ref{M1}) and  (\ref{M2}), that $\mathbb{A}^u$,
$\mathbb{A}^r$ and $\mathbb{A}^A$ are Hermitian, i.e.,
\begin{equation}
\overline{\mathbb{A}}{}^\mu{}^T={\mathbb{A}}^\mu\,.
\end{equation}
Similarly, the combination
$\mathbb{A}^\mu(n_\mu+l_\mu)$ can be seen to be positive definite (at
least) in a sufficiently small open neighbourhood of
${\widetilde{\mathcal{H}}_1}$ since  
\begin{equation}
\mathbb{A}^\mu(n_\mu+l_\mu)|_{\widetilde{\mathcal{H}}_1}=
  (\mathbb{A}^u+\mathbb{A}^r)|_{\widetilde{\mathcal{H}}_1}= 
  \mathbb{A}^{\rm{D}}+\mathbb{A}^{\Delta} 
\end{equation}  
thereby its determinant,
$det\left(\mathbb{A}^\mu(n_\mu+l_\mu)\right)$, takes the value $8$ on
${\widetilde{\mathcal{H}}_1}$.

\medskip

As a direct conclusion of all above, the
system comprised by (FR.1)-(FR.18) is a quasilinear symmetric
hyperbolic system for which the existence and uniqueness of solutions
is guaranteed which, in turn, justifies then the following
theorem  (see Theorem 2.\,of \cite{friedrich}).
\begin{theorem}\label{HF2} 
In the characteristic initial value problem to any `reduced initial
data set' there always exists a unique  solution to the vacuum
Einstein's equations.
\end{theorem}

\subsection{Further geometrical properties}  

In returning to our basic problem notice first that the distorted
black hole spacetimes under considerations are definitely not the most
generic configurations to which the above recalled results are known
to apply. Thereby, even a ``reduced initial data set'' should further
simplify. To see that this really happens let us recall first that
since the event horizon was supposed to be a Killing horizon the
bifurcate horizon $\widetilde{\mathcal{H}}^*$ is necessarily expansion
and shear free. This, in virtue of some of the results of \cite{r1}
(see Remarks 3.1 and 6.1 of that reference), implies that in a
distorted black hole spacetime with matter satisfying the dominant
energy condition, the spin coefficients $\lambda$ and $\mu$ vanish on
$\widetilde{\mathcal{H}}_1$, while $\sigma$ and $\rho$ were shown to
be identically zero on $\widetilde{\mathcal{H}}_2$.  It also
follows that the horizon Killing vector field ${\mathfrak{K}^*}^a$
is a repeated principal null vector of the Weyl and Ricci tensors on
$\widetilde{\mathcal{H}}^*=\widetilde{\mathcal{H}}_1\cup
\widetilde{\mathcal{H}}_2$. More precisely, it was shown in \cite{r1}
that the Ricci spinor components $\Phi _{22}$ and $\Phi _{21}$, as
well as, the Weyl spinor components $\Psi_3$ and $\Psi_4$ vanish on
$\widetilde{\mathcal{H}}_1$. Similarly, $\Phi _{00}$ and $\Phi _{01}$,
as well as, $\Psi_0$ and $\Psi_1$, vanish on
$\widetilde{\mathcal{H}}_2$.

\medskip

In addition to the gauge choices that have already been made, as a
consequence of the fact that  $\mathfrak{K}^a$ is the horizon  Killing
vector field, we also have.

\begin{lemma}\label{tau} 
The spin coefficient $\tau$ vanishes on $\widetilde{\mathcal{H}}_1$.
\end{lemma} 
  
\noindent\textbf{Proof}{\ } It follows from the definition
 $\tau$, along with the facts that $l_a=\nabla _a u$ and $l^an_a=1$ in
 $\widetilde{\mathcal{O}}$, that 
\begin{equation}\label{tau1} 
\tau=n^am^b\nabla _al_b=n^am^b\nabla _bl_a=-l^am^b\nabla _bn_a
\end{equation}
everywhere in $\widetilde{\mathcal{O}}$. Since $m^a$ is tangential to 
$\widetilde{\mathcal{N}}$,
$m^a|_{\widetilde{\mathcal{N}}}=\xi^A\partial_{x^A}$,  in evaluating
the term $m^b\nabla _bn_a$ on $\widetilde{\mathcal{N}}$ the way  $n^a$
extends from $\widetilde{\mathcal{N}}$ onto
$\widetilde{\mathcal{O}}_+=\{p\in \widetilde{\mathcal{O}}\,|\, u>0 \}$
does not matter. Thereby, in calculating  $m^b\nabla  _bn_a$ on
$\widetilde{\mathcal{N}}$ the substitution $n^a={1}/({\kappa_\circ
u})\cdot \mathfrak{K}^a$, see (\ref{kur}),   can be applied. Taking
into account, then, that $\pounds_m u=\pounds_l u=0$ everywhere in
$\widetilde{\mathcal{O}}$, moreover, that $\mathfrak{K}^a$ is a
Killing vector field, i.e. $\nabla_b \mathfrak{K}_a=-\nabla_a
\mathfrak{K}_b$ in $\widetilde{\mathcal{O}}_+$, it follows from
(\ref{tau1}) that
\begin{equation} 
\tau=-\frac{1}{\kappa_\circ u}
l^am^b\nabla_b\mathfrak{K}_a=\frac{1}{\kappa_\circ u} 
l^am^b\nabla _a \mathfrak{K}_b=l^am^b\nabla _a n_b
\end{equation}
on $\widetilde{\mathcal{N}}$.  This relation, along with the fact that
$n^a$ is parallel with respect to $l^a$, justifies then that  $\tau$
has to vanish on $\widetilde{\mathcal{N}}$. Due to the `wedge
reflection' symmetry of the extension $(\mathcal{O}^*,g_{ab}^*)$
$\tau$ also vanishes on
$\widetilde{\mathcal{H}}_1\setminus\widetilde{\mathcal{Z}}$ and, in
turn, by continuity, on the entire of
$\widetilde{\mathcal{H}}_1$. \hfill \fbox{}

\bigskip

Notice that the above lemma does not claim that the spin coefficient
$\tau$ should have to vanish for all the possible choices of a
complex null tetrad. It does  merely guarantee the vanishing of $\tau$
on $\widetilde{\mathcal{H}}_1$ for those tetrads which are
compatible with the gauge choices have been  made above.

\medskip

The following lemma provides a simple geometric characterisation of
the (future) event horizon $\mathcal{N}$ in terms of the spin
coefficient $\rho$. 

\begin{lemma}\label{rho} 
The $3$-parameter family of null geodesics with tangent $l^a$ is
non-contracting on the (future) event horizon $\widetilde{\mathcal{N}}$
in the direction of the domain of outer communication,
$\mathcal{D}_{\mathcal{N}}$, if and only if $\rho\geq 0$ at
$\widetilde{\mathcal{N}}$.
\end{lemma} 

\noindent\textbf{Proof}{\ } Start by recalling that the
spin-coefficient $\rho$ is defined as $\rho=m^a\overline{m}^b\nabla
_al_b$ which, along with the relations $\nabla _al_b=\nabla _bl_a$,
$g^{ab}=l^ak^b+k^al^b - m^a\overline{m}^b - 
\overline{m}^am^b$ and that all the tetrad vectors are parallelly
propagated with respect to $l^a$ as well as that $l^a$ is null
everywhere, gives that
\begin{equation}
\rho=m^a\overline{m}^b\nabla _al_b= -\frac12 g^{ab}\nabla _al_b
=-\frac12\theta_{(l)}=\frac12\theta_{(-l)}, 
\end{equation}
where $\theta_{(l)}$ denotes the expansion of the null congruence with
respect to $l^a$. Thus, the expansion of the  null geodesics geodesics
pointing towards $\mathcal{D}_{\mathcal{N}}$ on the (future) event
horizon 
$\widetilde{\mathcal{N}}$, with tangent vector 
$-l^a$, is  non-negative if and only if $\rho\geq 0$.  \hfill \fbox{}
\bigskip

\subsection{The determination of a full initial data set}\label{app} 

In virtue of Lemma\,\ref{lHF2} and Theorem\,\ref{HF2} from a reduced
initial data set the full information associated with a solution of
the vacuum Einstein's equations can be recovered in the  domain of
dependence of the initial data surface. Moreover, a reduced initial
data set, $\mathbb{V}^{red}_0$, is given as
\begin{equation} 
\mathbb{V}^{red}_0=\{\rho,\sigma,\mu,\lambda,\tau\,;
\,\xi^A\}|_{\widetilde{\mathcal{Z}}} \cup
\{\Psi_4\}|_{{\widetilde{\mathcal{H}}_1}} \cup 
\{\Psi_0\}|_{{\widetilde{\mathcal{H}}_2}} 
\end{equation}
Accordingly, in case of a considered distorted black hole spacetime,
we need to specify the spin-coefficients
$\rho,\sigma,\mu,\lambda,\tau$  and the vector field $\xi^A$ on
$\widetilde{\mathcal{Z}}$, moreover, Weyl spinor components $\Psi_4$
on ${\widetilde{\mathcal{H}}_1}$ and $\Psi_0$ on
${\widetilde{\mathcal{H}}_2}$.

\medskip

In virtue of the observations made in the previous subsection, the
reduced  initial data set simplifies considerably in case of our
configurations since  the initial data surface is comprised by two
expansion and shear free null geodesic congruences. This, along with
lemma\,\ref{tau}, implies that the only non-trivial quantity which can
``yet'' be freely specified as our initial data on
${\widetilde{\mathcal{H}}_1}\cup {\widetilde{\mathcal{H}}_2}$ is
nothing but the vector field $\xi^A$ on $\widetilde{\mathcal{Z}}$.

\medskip

To illustrate the way a full initial data set is produced from a
reduced one, moreover, to appreciate the robustness of the setup of
\cite{newman:penrose,friedrich} we shall carry out the determination
of a full initial data set $\mathbb{V}_0$, on
${\widetilde{\mathcal{H}}_1} \cup {\widetilde{\mathcal{H}}_2}$, from
the significantly simplified one
\begin{equation} 
\mathbb{V}^{red}_0=\{\rho=\sigma=\mu=\lambda=0\,;
\,\xi^A\}|_{\widetilde{\mathcal{Z}}} \cup
\{\Psi_4=\tau=\nu=\gamma=0\}|_{{\widetilde{\mathcal{H}}_1}} \cup 
\{\Psi_0=0\}|_{{\widetilde{\mathcal{H}}_2}}, 
\end{equation}
which is compatible with all of our former observations concerning the
geometrical properties of a bifurcate Killing horizon of a stationary
distorted black hole, as well as, with all of the gauge conditions we
have made.

\medskip

Let us start by considering the ``inner equations'' we have on
$\widetilde{\mathcal{Z}}$. Notice first that (NP.6.11k) and
(NP.6.11m) immediately implies that both $\Psi_1$ and $\Psi_3$
vanishes on $\widetilde{\mathcal{Z}}$.  Furthermore, (NP.6.10f), along
with our gauge conditions $\tau=\overline{\alpha}+\beta$ in
$\widetilde{\mathcal{O}}$ and, in particular, $\tau=0$ on
${\widetilde{\mathcal{H}}_1}$, 
gives that 
\begin{equation} 
\delta \overline{\xi}^A - \overline{\delta} \xi^A=
2\,(\overline{\beta}\,\xi^A - \beta\,\overline{\xi}^A). 
\label{beta} 
\end{equation}
This equation can be solved algebraically for $\beta$ and
$\overline{\beta}=-\alpha$ on $\widetilde{\mathcal{Z}}$. By applying
then (NP.6.11l) we immediately get
\begin{equation} 
-\delta \overline{\beta} - \overline{\delta} \beta=
4\,\beta\,\overline{\beta}-\Psi_2. 
\label{psi2} 
\end{equation}
which fixes the value of $\Psi_2$ on $\widetilde{\mathcal{Z}}$. Notice
that the last relation also imply that $\Psi_2$ is necessarily real on
$\widetilde{\mathcal{Z}}$. 

Consider now the inner equations on ${\widetilde{\mathcal{H}}_2}$.
First of all, since $\Psi_0\equiv 0$ there (NP.6.12a), along with the
fact that $\Psi_1|_{\widetilde{\mathcal{Z}}}\equiv 0$, implies that
$\Psi_1\equiv 0$ on ${\widetilde{\mathcal{H}}_2}$. Similarly,  since
$\rho|_{\widetilde{\mathcal{Z}}}\equiv 0$ and
$\sigma|_{\widetilde{\mathcal{Z}}}\equiv 0$, (NP.6.11a) and (NP.6.11b)
imply that $\rho\equiv 0$ and $\sigma\equiv 0$ on
${\widetilde{\mathcal{H}}_2}$. The vanishing of $\rho$, $\sigma$ and
$\Psi_1$ on ${\widetilde{\mathcal{H}}_2}$ can then be used, 
along with (NP.611c-d-e), to conclude that 
\begin{equation} 
\mathrm{D}\alpha=\mathrm{D}\beta=\mathrm{D}\tau=0
\label{dD} 
\end{equation}
on ${\widetilde{\mathcal{H}}_2}$, which along with the vanishing of
$\tau$ on ${\widetilde{\mathcal{Z}}}$ does imply that $\tau\equiv 0$ on
${\widetilde{\mathcal{H}}_2}$. Similarly, (NP.6.12b) gives then 
\begin{equation}\label{psi2n2}
\mathrm{D}\Psi_2=0
\end{equation} 
on ${\widetilde{\mathcal{H}}_2}$. In virtue of (NP.6.11g) and
(NP.6.11i) we also have then that $\lambda\equiv 0$ on
${\widetilde{\mathcal{H}}_2}$ since $\lambda$ vanishes on
$\widetilde{\mathcal{Z}}$. Two other spin coefficients, 
$\gamma$ and $\mu$, can be determined with the help of (NP.6.11f) and
(NP.6.11h) which, along with (\ref{psi2n2}) and their vanishing on
$\widetilde{\mathcal{Z}}$, give that $\gamma=r\cdot \Psi_2$ and
$\mu=r\cdot \Psi_2$ on ${\widetilde{\mathcal{H}}_2}$. 

\medskip

By completely analogous reasoning the inner equations on
${\widetilde{\mathcal{H}}_1}$, along with the vanishing of $\nu,
\gamma$ and $\tau$ there, which follow from our gauge choice, can be
used to justify the followings. First, since $\Psi_4\equiv 0$ there
(NP.6.12h), along with the fact that
$\Psi_3|_{\widetilde{\mathcal{Z}}}\equiv 0$, implies that
$\Psi_3\equiv 0$ on ${\widetilde{\mathcal{H}}_1}$. Similarly,  since
$\mu|_{\widetilde{\mathcal{Z}}}\equiv 0$ and
$\lambda|_{\widetilde{\mathcal{Z}}}\equiv 0$, (NP.6.11n) and
(NP.6.11j) imply that $\mu\equiv 0$ and $\lambda\equiv 0$ on
${\widetilde{\mathcal{H}}_1}$. The vanishing of $\mu$, $\lambda$ and
$\Psi_3$ on ${\widetilde{\mathcal{H}}_1}$ can be used,  along with
(NP.611r-o-p), to conclude that
\begin{equation} 
\Delta \alpha=\Delta \beta=\Delta \sigma =0
\label{dDen1} 
\end{equation}
on ${\widetilde{\mathcal{H}}_1}$. This latter relation, along with the
vanishing of $\sigma$ on ${\widetilde{\mathcal{Z}}}$, implies
that $\sigma\equiv 0$ on
${\widetilde{\mathcal{H}}_1}$. Similarly, (NP.6.12g) gives then 
\begin{equation}\label{psi2n1}
\Delta\Psi_2=0
\end{equation} 
on ${\widetilde{\mathcal{H}}_1}$.  The only remaining non-trivial spin
coefficient is $\rho$ which is determined with the help of (NP.6.11q),
along with (\ref{psi2n1}) and its vanishing on
$\widetilde{\mathcal{Z}}$, as $\rho=- u\cdot \Psi_2$ on
${\widetilde{\mathcal{H}}_1}$.
 
To have a full initial data set $\mathbb{V}_0$ on
${\widetilde{\mathcal{H}}_1} \cup {\widetilde{\mathcal{H}}_2}$, in
addition to what we have already derived above, we also need to
determine the behaving all of the Weyl spinor components on
${\widetilde{\mathcal{H}}_1} \cup {\widetilde{\mathcal{H}}_2}$, as
well as, the value of $\nu$ on ${\widetilde{\mathcal{H}}_2}$. For
instance, from (NP.6.12f) we get that $\Delta \Psi_1-\delta\Psi_2=0$
on ${\widetilde{\mathcal{H}}_1}$, which, in virtue of
${\Psi_1}|_{\widetilde{\mathcal{Z}}}=0$ and the $u$-independentness of
$\Psi_2$ implies that
\begin{equation}
\Psi_1=u\cdot\delta\Psi_2
\end{equation} 
on ${\widetilde{\mathcal{H}}_1}$. By an analogous argument, we also
get, from (NP.6.12e) and from what we have just established, that
\begin{equation}
\Psi_0=\frac{1}{2}u^2\left(\delta^2\Psi_2-2\beta\cdot\delta\Psi_2\right) 
\end{equation} 
holds on ${\widetilde{\mathcal{H}}_1}$.

Completely parallel to the reasoning applied in the previous paragraph
we can also show, by making use of (NP.6.12c) and (NP.6.12d), that 
\begin{equation}\label{psi3n2}
\Psi_3=r\cdot\overline{\delta}\Psi_2
\end{equation} 
and 
\begin{equation}
\Psi_4=\frac{1}{2}r^2\left(\overline{\delta}^2\Psi_2
+2\alpha\cdot\overline{\delta}\Psi_2\right)    
\end{equation}
hold on ${\widetilde{\mathcal{H}}_2}$. Finally, by making use of
(NP.6.11i) and (\ref{psi3n2}) the value of $\nu$ can
be determined on ${\widetilde{\mathcal{H}}_2}$ as 
\begin{equation}
\nu=\frac{1}{2}r^2\cdot\overline{\delta}\Psi_2.
\end{equation}

\bigskip

It is also informative to collect what we have already established on
${\widetilde{\mathcal{H}}_1} \cup {\widetilde{\mathcal{H}}_2}$ (see
Table\,\ref{table:data}).  In order to simplify some of the involved
expressions the ``{\it edth}''-operator, $\eth$,  of Newman
and Penrose \cite{newman66:_note_bondi_metzn_sachs} has been applied.
For instance, since $\Psi_2$ is a $\{0,0\}$-type scalar
$\delta\Psi_2$ can also be written as
\begin{equation}\label{rel1}
\delta\Psi_2=\eth\Psi_2\,.
\end{equation}
Noticing also that the $\{p,q\}$-type of
$\delta\Psi_2$ is $\{1,-1\}$---which is in accordance with the fact
that $\delta\Psi_2$ possesses ``{\it spin-weight}\,''
$s=\frac12(p-q)=1$ and ``{\it boost-weight}\,''
$b=\frac12(p+q)=0$---we get, in virtue of (2.14) of \cite{GHP} and
by the relation $\tau=\overline{\alpha}+\beta=0$ which holds on
$\mathcal{Z}$, that
\begin{equation}\label{rel2}
\delta^2\Psi_2-2\beta\cdot\delta\Psi_2=\eth^2\Psi_2. 
\end{equation}
Applying then (\ref{rel1}) and (\ref{rel2}), along with their complex
conjugates, the full initial data set can be given as in
Table\,\ref{table:data}.
\begin{table}[h!]
\centering 
\begin{tabular}{|c|c|c|} 
\hline $\phantom{\frac{\frac12}{\frac12}{4}}$ $
 {\widetilde{\mathcal{H}}_1}$ &  ${\widetilde{\mathcal{Z}}}$ &
 ${\widetilde{\mathcal{H}}_2}$ \\ \hline \hline
 $\phantom{\frac{\frac12}{\frac12}{4}}$ $\rho = -u \cdot \Psi_2$ &
 $\rho =0$ & $\rho =0$ \\  \hline
 $\phantom{\frac{\frac12}{\frac12}{4}}$  $\mu = 0$ & $\mu =0$ & $\mu =
 r \cdot \Psi_2$ \\ \hline
 $\phantom{\frac{\frac12}{\frac12}{\frac12}{4}}$
 $\sigma=\lambda=\tau=0$ & $\sigma=\lambda=\tau=0$ &
 $\sigma=\lambda=\tau=0$ \\ \hline
 $\phantom{\frac{\frac12}{\frac12}{4}}$ $\Delta \alpha=\Delta \beta=0$
 & $\alpha,\beta:\ \tau=\overline{\alpha}+\beta=0$  &
 $\mathrm{D}\alpha=\mathrm{D}\beta=0$ \\ \hline
 $\phantom{\frac{\frac12}{\frac12}{4}}$ $\Delta\Psi_2=0$ & $\xi^A\ \&
 \ \alpha,\beta \rightarrow \Psi_2$  & $\mathrm{D} \Psi_2 =0$ \\
 \hline $\phantom{\frac{\frac12}{\frac12}{4}}$
 $\Psi_0=\frac{1}{2}u^2\cdot{\eth}{}^2\Psi_2$
 & $\Psi_0=0$ & $\Psi_0=0$ \\ \hline
 $\phantom{\frac{\frac12}{\frac12}{4}}$  $\Psi_1=u\cdot\eth\Psi_2$ &
 $\Psi_1=0$ & $\Psi_1=0$ \\ \hline
 $\phantom{\frac{\frac12}{\frac12}{4}}$  $\Psi_3=0$ & $\Psi_3=0$ &
 $\Psi_3=r\cdot\overline{\eth}\Psi_2$ \\ \hline
 $\phantom{\frac{\frac12}{\frac12}{4}}$ $\Psi_4=0$ & $\Psi_4=0$ &
 $\Psi_4=\frac{1}{2}r^2\cdot\overline{\eth}{}^2\Psi_2$ \\ \hline
 $\phantom{\frac{\frac12}{\frac12}{4}}$  (gauge) $\nu=\gamma=0$\ \ \
 $\rightarrow$ & $\nu=\gamma=0$ \ \ \ $\rightarrow$ & \ \
 $\nu=\frac{1}{2}r^2\cdot\overline{\eth}\Psi_2$,  $\gamma=r \cdot
 \Psi_2$  \ \   \\ \hline
\end{tabular}
\caption{\small The full initial data set $\mathbb{V}_0$, on the
  intersecting null hypersurfaces ${\widetilde{\mathcal{H}}_1} \cup
  {\widetilde{\mathcal{H}}_2}$. }\label{table:data}
\end{table}

\subsection{The behaviour of the curvature along the generators of
  $\mathcal{N}$} 

This short subsection is to show that, in general, ``{\it parallelly
propagated}\,'' curvature singularities may occur along the
generators of $\mathcal{H}^*$.  To get some hints about this point it
is rewarding to inspect Table\,\ref{table:data} for a short
while. What might not be too striking for the first glance is the
$u$-dependence of the Weyl spinor components $\Psi_0$ and $\Psi_1$
along the null geodesic generators of ${\widetilde{\mathcal{H}}_1}$,
and similarly, the $r$-dependence of $\Psi_3$ and $\Psi_4$ along the
null geodesic generators of ${\widetilde{\mathcal{H}}_2}$. Obviously,
all of these quantities vanish at the bifurcation surface but when
we approach the asymptotic ends of ${\widetilde{\mathcal{H}}_1}$ or
${\widetilde{\mathcal{H}}_2}$ they blow up.

\medskip

Clearly, such a blow up of the Weyl spinor components does not occur
if $\delta\Psi_2=\eth\Psi_2=0$ on  $\mathcal{Z}$ but this condition
seems to be very restrictive. It is not hard to justify that it is
equivalent to requiring $\mathcal{Z}$ to be a metric sphere, and this
condition in the pure vacuum case, considered in this section, is
satisfied only by the Schwarzschild solution. Therefore the following
question immediately emerge. Is it true that all the other black hole
spacetimes, including the Kerr solution, possess the indicated type of
curvature blow up? It is also important to know whether we do really
have a blow up of certain measurable quantities or what is indicated
above is simply  the consequence of an inappropriate choice of a
frame field along the null geodesic generators of the bifurcate event
horizon. In this respect it is informative to have the following.
\begin{proposition}\label{ndl} 
The blow up of the Weyl
spinor component $\Psi_1$ along the generators of
${\widetilde{\mathcal{H}}_1}$ is 
always associated with true ``{\it parallelly
propagated}\,'' curvature singularity.
\end{proposition} 
\noindent\textbf{Proof}{\ }  The validity of the above assertion can
be justified by the inspection of components of the Weyl
tensor, with respect to  basis fields parallelly propagated along the
generators of ${\widetilde{\mathcal{H}}_1}$. First we shall show that
unit norm real vector fields $X^a_{^{(A)}}$, $A=3,4$, can be chosen so
that $\{l^a,n^a,X^a_{^{(A)}}\}$ will serve as a suitable
pseudo-orthonormal parallelly propagated basis fields along the
generators.

\medskip 

To see this notice first that $n^a$ is parallelly propagated with
respect to itself on ${\widetilde{\mathcal{H}}_1}$. Next it will be
verified that $l^a$ is also parallelly propagated with respect to $n^a$
on ${\widetilde{\mathcal{H}}_1}$. In doing so notice that
$n^e\nabla_e l^a=0$  on ${\widetilde{\mathcal{H}}_1}$ if all the
contractions $n_an^e\nabla_e l^a$, $l_an^e\nabla_e l^a$ and
$m_an^e\nabla_e l^a$ vanish there. The first contraction vanishes
since $n_an^e\nabla_e l^a=-l_an^e\nabla_e n^a$ and $n^a$ is parallelly
propagated with respect to itself on
${\widetilde{\mathcal{H}}_1}$. The second contraction is zero because
$l^a$ is null everywhere. Finally, the vanishing of the third
contraction on  ${\widetilde{\mathcal{H}}_1}$ is guaranteed by
Lemma\,\ref{tau}.

\medskip 

To get the desired basis field $\{l^a,n^a,X^a_{^{(A)}}\}$  the
spacelike unit vector fields $X^a_{^{(A)}}$, $A=3,4$, are also needed
to be specified.  In this respect it is important to note that by
choosing the smooth functions $\Theta_{^{(A)}}:h_1\rightarrow
\mathbb{R}$ along any generator $h_1$ of ${\widetilde{\mathcal{H}}_1}$
suitably  the desired unit norm real spacelike vector fields
$X^a_{^{(A)}}$ can be given as the linear combinations
$X^a_{^{(A)}}=\cos\Theta_{^{(A)}}\cdot x^a+\sin\Theta_{^{(A)}}\cdot
y^a$ of the unit spacelike vectors $x^a=\frac{1}{\sqrt{2}}
(m^a+\overline{m}^a)$ and $y^a=\frac{i}{\sqrt{2}}
(m^a-\overline{m}^a)$.  Then by making use of the definition
\begin{equation}
\Psi_1=-C_{abcd}l^an^bl^cm^d\,,
\end{equation} 
where $C_{abcd}$ denotes the Weyl tensor, it is straightforward to
verify that 
\begin{equation}\label{PAX}
C_{abcd}l^an^bl^c X^d_{^{(D)}}=-\sqrt{2}\cdot\left[
  \cos\Theta_{^{(D)}}\cdot  
  \mathfrak{Re}(\Psi_1) - \sin\Theta_{^{(D)}}\cdot
  \mathfrak{I m}(\Psi_1) 
\right] 
\end{equation} 
Notice also that in our case, since $X^a_{^{(3)}}$ and $X^a_{^{(4)}}$ are
spacelike members of the pseudo-orthonormal tetrad
$\{l^a,n^a,X^a_{^{(A)}}\}$ they should also be orthogonal to each
other. Thereby, we may assume, without loss of generality, that
$\Theta_{^{(4)}}$ is chosen so that
$\Theta_{^{(4)}}=\Theta_{^{(3)}}+\pi/2$ everywhere along the generator
$h_1$. Then, (\ref{PAX}), along with this relation, can be used to
justify
\begin{equation}\label{PAX2}
\left(C_{abcd}l^an^bl^c X^d_{^{(3)}}\right)^2+
\left(C_{abcd}l^an^bl^c X^d_{^{(4)}}\right)^2=2\left[
\left(\mathfrak{Re}(\Psi_1)\right)^2+
\left(\mathfrak{Im}(\Psi_1)\right)^2 
 \right]\,,
\end{equation} 
which immediately implies that whenever the Weyl spinor component
$\Psi_1$ blows up along $h_1$ while $u\rightarrow \infty$ then either
of the contractions $C_{abcd}l^an^bl^c X^d_{^{(A)}}$, $A=3,4$,  must
also blow up there.  \hfill\fbox{}\bigskip

We would like to recall that the possible appearance of ``parallelly
propagated'' curvature singularities in black hole spacetimes the
event horizon of which is also Killing horizon was already mentioned
in \cite{r1} (see Remark 6.2 there for more details).  
Notice also that the existence of the indicated $p.p.$ curvature
singularity is not associated with the incompleteness of the null
geodesic generators of the bifurcate horizon. It is also interesting
that there is a difference between the possible strengths of the
associated curvature blows up. While $\Psi_1$ and $\Psi_3$ might blow
up only linearly when a blow up occurs in case of $\Psi_0$ and
$\Psi_4$ that has to be quadratic with respect to the associated
synchronised affine parameters.

\medskip

Finally, we would like to emphasise that the above discussed $p.p.$
curvature singularities are not strong enough to be ``{\it scalar
curvature singularity}\,'', i.e., neither of the scalar invariants of
the Weyl tensor blows up along the generators of the event
horizon. Thereby, the existence of these $p.p.$ curvature
singularities does merely indicate that in general certain {\it
tidal-force} effects increase ``in time'', along the black hole  event
horizon.

\section{The local existence and uniqueness results}\label{key}
\renewcommand{\theequation}{\ref{key}.\arabic{equation}}
\setcounter{equation}{0}

As a direct consequence of the preceding subsections,
Subsections\,\ref{nchfr}\,-\,\ref{app}, we immediately have that in
the domain of dependence of the initial data surface $\mathcal{H}_1
\cup \mathcal{H}_2$---which is limited to a part of the ``black'' and
``white'' hole regions in $\mathcal{O}^*$, i.e. to a part of the
causal future, $J^+[\mathcal{Z}]\cap\mathcal{O}^*$, and past,
$J^-[\mathcal{Z}]\cap\mathcal{O}^*$, of $\mathcal{Z}$,
respectively---the solution to the reduced vacuum equations is
uniquely determined once the vector field $\xi^A$, or equivalently
the $2$-metric $g^{AB}$, is specified on $\mathcal{Z}$. Accordingly,
we have then the following.

\begin{theorem}\label{IR}
Consider a vacuum spacetime $(M,g_{ab})$ of type B and with a
non-degenerate (future) event horizon ${\mathcal{N}}$. Then the
spacetime metric $g_{ab}$ is uniquely determined in the black hole
region once the $2$-metric of the space of Killing orbits on
${\mathcal{N}}$ is fixed.
\end{theorem} 

This section is to identify those conditions that may guarantee that
the geometry of the considered distorted ``stationary'' vacuum black
hole spacetimes gets to be uniquely determined  also on the domain of
outer communication side.  In doing so the fact that
\begin{equation}
\mathfrak{K}^a=\kappa_{\circ}u k^a
\end{equation}
is a horizon Killing vector field on $(M,g_{ab})$ will play an
important role.

\medskip 

To start off consider first the effect of the null rotation
\begin{equation}\label{nrot}
\hat l^a=A^{-1}l^a,\ \ \ \hat n^a=A n^a,\ \ \ \hat m^a=m^a,
\end{equation}
which leaves the directions of $l^a$ and $n^a$ fixed and is, in fact,
nothing but the boost transformation in the $l-n$ plane with
\begin{equation}
A=\kappa_{\circ}u
\end{equation}
on the part ${\widetilde{\mathcal{O}}_+}$ of
${\widetilde{\mathcal{O}}}$ with $u>0$. Then, the transformed
spin-coefficients, the Weyl-spinor components and the derivatives can
be related to the original ones as
\begin{equation}\label{trans}
\begin{array}{llllll} 
\hat{\kappa}=\hat{\epsilon}=\hat{\pi}=0 & \hat{\rho}=A^{-1}\rho &
  \hat{\alpha}= \alpha & \hat{\gamma}= 
  A\gamma-\frac12\kappa_{\circ} & & \hat\Psi_0=A^{-2} \Psi_0 \\
& \hat{\sigma}=A^{-1}\sigma & \hat{\beta}= \beta & \hat{\lambda}=
  A\lambda & & \hat\Psi_1=A^{-1} \Psi_1 \\
& & \hat{\tau}= \tau & \hat{\mu}=  A\mu & & \hat\Psi_2= \Psi_2\\
& & & \hat{\nu}= A^2\nu & & \hat\Psi_3=A \Psi_3 \\
& & & & & \hat\Psi_4=A^2 \Psi_4
\end{array}
\end{equation}
\begin{equation}\label{der} 
\hat \mathrm{D}=A^{-1}\mathrm{D}
,\ \ \ 
\hat \Delta=A \Delta
,\ \ \ 
\hat \delta=\delta\,.
\end{equation}
By making use the coordinate transformation 
\begin{equation} 
\hat u=\frac{1}{\kappa_{\circ}}\ln u
,\ \ \ 
\hat r=\kappa_{\circ} u r 
,\ \ \ 
\hat x^a=x^A
\end{equation}
the coefficients of the coordinate components of the ``hatted'' tetrad
fields  
\begin{equation} 
\hat l^\mu (=\mathfrak{L}^\mu) =\delta ^\mu {}_{\hat r},\ \ \hat n^\mu
     =\delta ^\mu 
     {}_{\hat u}+\hat U \delta ^\mu 
{}_{\hat r}+\hat X ^A\delta ^\mu {}_A, \ \ \hat m^\mu =\hat{\omega}
     \delta ^\mu {}_{\hat r}+\hat{\xi}^A\delta 
^\mu {}_A  \label{tet2} 
\end{equation}
can be seen to take the form 
\begin{equation} 
\hat{\omega}=A {\omega},\ \ \ \hat{\xi}^A={\xi}^A,\ \ \ \hat X^A=X^A,
\ \ \ \hat U= A^2 U + \kappa_{\circ} r A\,.
\end{equation}
The covariance of the Newman-Penrose equations, i.e., the fact that
they necessarily possess the same form in terms of the ``hatted'' variables
as they had before, can be justified by a direct calculation simply by
substituting all of these relations to the original Newman-Penrose
equations. 

In addition, it is also straightforward to see that the argument
presented in Subsection\,\ref{nchfr} does apply to the ``hatted''
version of the reduced vacuum field equations,  (FR.1)-(FR.18), i.e.,
these ``hatted'' equations do give rise to a determined system of
the form
\begin{equation}\label{eqVh}
\hat \mathbb{A}^{\hat \mu} \cdot \partial_{\hat \mu} \hat \mathbb{V} +
\hat\mathbb{B}=0, 
\end{equation}
where the matrices $\hat \mathbb{A}^{\hat \mu}$ and $\hat\mathbb{B}$,
besides depending smoothly on the vector valued variable 
\begin{equation}\label{hatV} 
\hat\mathbb{V}=(\hat\xi^A,\hat\omega,\hat X^A,\hat U;\hat\rho,\hat\sigma,
\hat\tau,\hat\alpha,\hat\beta,\hat\gamma,\hat\lambda, 
\hat\mu,\hat\nu;\hat\Psi_0,\hat\Psi_1, \hat\Psi_2,\hat\Psi_3,\hat\Psi_4)\,,
\end{equation}
along with its complex conjugate $\overline{\hat\mathbb{V}}$, are so
that for any value of ${\hat \mu}$ the matrices
$\hat\mathbb{A}^{\hat\mu}$ are Hermitian, i.e.,
$\overline{\hat\mathbb{A}}{}^{\hat\mu}{}^T={\hat\mathbb{A}}^{\hat\mu}$
and the combination $\hat\mathbb{A}^{\hat\mu}(\hat n_{\hat\mu}+\hat
l_{\hat\mu})$ is positive definite (at least) in a sufficiently small
neighbourhood of ${\widetilde{\mathcal{N}}}$ in
${\widetilde{\mathcal{O}}_+}$. Accordingly, (\ref{eqVh}) can be seen
to be a quasilinear symmetric  hyperbolic system for the variable
$\hat\mathbb{V}$.

This property, along with the fact that $\mathfrak{K}^a= \left(\partial/
\partial \hat u \right)^a$ is a Killing vector field on
${\widetilde{\mathcal{O}}_+}$ will play important role in the rest of
this section. Recall that, in the vacuum case, the vanishing of the
Lie derivative of the Weyl tensor, $\pounds_{\mathfrak{K}} C_{abcd}$,
immediately follows from the fact that $\mathfrak{K}^a$ is a Killing
vector field. This can be used in justifying the claim of the following.

\begin{lemma}\label{LPsi01}
Assume as above that ${\widetilde{\mathcal{O}}_+}$ is the part of
${\widetilde{\mathcal{O}}}$ which can be covered by the Gaussian null
coordinates $(\hat u,\hat r,\hat x^3,\hat x^4)$. Then, 
\begin{equation}\label{LPsi}
\pounds_{\mathfrak{K}}\hat\Psi_0=-C_{abcd}\hat l^a(\partial_{\hat x^B})^b
\hat l^c (\partial_{\hat  x^D})^d \left[\pounds_{\mathfrak{K}}
  \hat{\xi}^B \hat{\xi}^D + \hat{\xi}^B \pounds_{\mathfrak{K}}
  \hat{\xi}^D \right]
\end{equation}
in ${\widetilde{\mathcal{O}}_+}$.
\end{lemma} 

\noindent\textbf{Proof}{\ } 
We have by the definition of
$\hat\Psi_0$  that
\begin{equation}
\hat\Psi_0=-C_{abcd}\hat l^a\hat m^b
\hat l^c \hat m^d
\end{equation}
which, along with the symmetries of $C_{abcd}$ and the third relation
in (\ref{tet2}), gives that  
\begin{equation}
\hat\Psi_0=-C_{abcd}\hat l^a(\partial_{\hat x^B})^b
\hat l^c (\partial_{\hat  x^D})^d \hat{\xi}^B \hat{\xi}^D. 
\end{equation} 
The statement of our lemma follows then from the facts that
$\pounds_{\mathfrak{K}} C_{abcd}$ vanishes, moreover, that 
$\mathfrak{K}=\partial_{\hat  u}, \hat l=\partial_{\hat  r},
\partial_{\hat  x^3}$ and $\partial_{\hat  x^4 }$ are coordinate basis
vector fields so they commute.
\hfill \fbox{} \bigskip 

\begin{lemma}\label{LPsi02}
The derivatives $\pounds_{\mathfrak{K}}\hat{\xi}^A$,
$\pounds_{\mathfrak{K}}\hat\rho$, $\pounds_{\mathfrak{K}}\hat\sigma$
and $\pounds_{\mathfrak{K}}\hat\Psi_0$ vanish on
${\widetilde{\mathcal{O}}_+}$.   
\end{lemma} 

\noindent\textbf{Proof}{\ } Notice first that, in virtue of
Table\,\ref{table:data} and by (\ref{trans}),  $\hat\xi^A=\xi^A$,
$\hat\rho=-\Psi_2/\kappa_{\circ}$ and  $\hat\sigma=0$ all are
independent of $\hat u$ on ${\widetilde{\mathcal{N}}}$ whence the
deviates $\pounds_{\mathfrak{K}}\hat{\xi}^A$,
$\pounds_{\mathfrak{K}}\hat\rho$, $\pounds_{\mathfrak{K}}\hat\sigma$
vanish there. In addition, since $\mathfrak{K}=\partial_{\hat  u}$ and
$\hat l=\partial_{\hat  r}$ commute we also have, in virtue of
(NP.6.10a), (NP.6.11a) and (NP.6.11b), that
\begin{eqnarray}   
\hat \mathrm{D}(\pounds_{\mathfrak{K}}\hat{\xi}^A)&=&
  \pounds_{\mathfrak{K}}\hat\rho\cdot\hat{\xi}^A 
  +\hat\rho\cdot\pounds_{\mathfrak{K}}\hat{\xi}^A
  + \pounds_{\mathfrak{K}}\hat\sigma\cdot{\overline{\hat{\xi}}}{}^A 
  +\hat\sigma\cdot\pounds_{\mathfrak{K}}{\overline{\hat{\xi}}}{}^A
\label{s1}\\
\hat \mathrm{D}(\pounds_{\mathfrak{K}}\hat\rho)&=&
  2\hat\rho\cdot\pounds_{\mathfrak{K}}\hat\rho 
  +\pounds_{\mathfrak{K}}\hat\sigma\cdot{\overline{\hat{\sigma}}}
  +\hat\sigma\cdot\pounds_{\mathfrak{K}}{\overline{\hat{\sigma}}}
\label{s2}\\
\hat \mathrm{D}(\pounds_{\mathfrak{K}}\hat\sigma)&=&
 2\pounds_{\mathfrak{K}}\hat\rho\cdot\hat\sigma 
 +2\hat\rho\cdot\pounds_{\mathfrak{K}}\hat\sigma 
+\pounds_{\mathfrak{K}}\hat\Psi_0\,. \label{s3}
\end{eqnarray}
Notice also that, in virtue of Lemma\,\ref{LPsi01}, the system
comprised by (\ref{s1}), (\ref{s2}) and (\ref{s3}),  along with the
complex conjugate of (\ref{s1}) and (\ref{s3}), gives rise to be a
homogeneous linear system for the variables
$\pounds_{\mathfrak{K}}\hat{\xi}^A$, $\pounds_{\mathfrak{K}}\hat\rho$,
$\pounds_{\mathfrak{K}}\hat\sigma$,
$\pounds_{\mathfrak{K}}\overline{\hat{\xi}}{}^A$,
$\pounds_{\mathfrak{K}}\overline{\hat\sigma}$, along the null geodesic
congruences with tangent $\hat l^a$ in ${\widetilde{\mathcal{O}}_+}$.
This, along with the vanishing of the deviates
$\pounds_{\mathfrak{K}}\hat{\xi}^A$, $\pounds_{\mathfrak{K}}\hat\rho$,
$\pounds_{\mathfrak{K}}\hat\sigma$ on ${\widetilde{\mathcal{N}}}$,
justifies then the major part of our claim. What remains to be shown
is the vanishing of $\pounds_{\mathfrak{K}}\hat\Psi_0$ in
${\widetilde{\mathcal{O}}_+}$ which follows then in virtue of
(\ref{LPsi}).     \hfill \fbox{}
\bigskip

The assertion of the following lemma can be justified by an argument
that is completely analogous to the one applied above. First noticing
that in virtue of Table\,\ref{table:data} and (\ref{trans}) the
$\pounds_{\mathfrak{K}}$-derivatives of quantities in question vanish
on ${\widetilde{\mathcal{N}}}$. Then, by applying the pertinent
Newman-Penrose equations a homogeneous linear system can be derived,
along the null geodesic congruences with tangent $\hat l^a$, for the
$\pounds_{\mathfrak{K}}$-derivatives of the considered quantities
which can, finally, be used to justify the vanishing of the relevant
$\pounds_{\mathfrak{K}}$-derivatives everywhere in
${\widetilde{\mathcal{O}}_+}$.

\begin{lemma}\label{LPsi03}
The Lie derivatives $\pounds_{\mathfrak{K}}\hat{X}^A$,
$\pounds_{\mathfrak{K}}\hat\omega$,
$\pounds_{\mathfrak{K}}\hat\alpha$, $\pounds_{\mathfrak{K}}\hat\beta$,
$\pounds_{\mathfrak{K}}\hat\gamma$, $\pounds_{\mathfrak{K}}\hat
\lambda$, $\pounds_{\mathfrak{K}}\hat\mu$,
$\pounds_{\mathfrak{K}}\hat\nu$, $\pounds_{\mathfrak{K}}\hat\Psi_1$,
$\pounds_{\mathfrak{K}}\hat\Psi_2$, $\pounds_{\mathfrak{K}}\hat\Psi_3$
and $\pounds_{\mathfrak{K}}\hat\Psi_4$  vanish on
${\widetilde{\mathcal{O}}_+}$.
\end{lemma} 

As a conclusion of the above lemmas we have that whenever
${\mathfrak{K}}^a=(\partial/\partial {\hat u})^a$ is a Killing vector
field then all the metric functions, the spin-coefficients and the
Weyl spinor components, appearing in the vector valued variable
(\ref{hatV}) have to be $\hat u$-independent in
${\widetilde{\mathcal{O}}_+}$. As an immediate implication we have
then that in ${\widetilde{\mathcal{O}}_+}$ the derivative operator
$\hat \Delta$ simplifies to
\begin{equation}
\hat \Delta=\hat U\cdot\hat \mathrm{D} +  \hat X ^A
  \partial_{\hat 
  x^A} \,. 
\end{equation}
Thereby, it seems to be plausible to consider---which will be done
hereafter---equation (\ref{eqVh}) as system of first order partial
differential  equations (PDEs) on null hypersurfaces intersecting
${\widetilde{\mathcal{N}}}$ transversely. To make this to be more
concrete, take any of the smooth cross-sections
${\widetilde{\mathcal{Z}}}_{\hat u}$ of
${\widetilde{\mathcal{N}}}$. Define ${\widetilde{\mathcal{N}}}_{\hat
u}^{T}$ to be the null hypersurface spanned by the null geodesics with
tangent $\hat l^a=\mathfrak{L}^a$ in ${\widetilde{\mathcal{O}}_+}$ and which
intersect ${\widetilde{\mathcal{N}}}$ at the points of
${\widetilde{\mathcal{Z}}}_{\hat u}$. Then, on
${\widetilde{\mathcal{N}}}_{\hat u}^{T}$ the reduced equations
(\ref{eqVh}) can be given as
\begin{equation}\label{eqV3h}
\hat\mathbb{A}^{\hat r} \cdot \partial_{\hat r} \hat\mathbb{V}+
\hat\mathbb{A}^{\hat A} \cdot \partial_{\hat A} \hat\mathbb{V} +
\hat\mathbb{B}=0, 
\end{equation}
with Hermitian coefficient matrices 
\begin{equation}\label{r}
\hat\mathbb{A}^{\hat r}=\left( {\begin{array}{c|c}
\mathbf{1} & \mathbf{0}\\ \hline 
\mathbf{0} &  \begin{array}{rrrrr}
\hat U  & -\hat\omega & 0 & 0 & 0 \\
 -\overline{\hat\omega} & 1 + \hat U  & -\hat\omega & 0 & 0 \\
 0 & -\overline{\hat\omega} & 1+ \hat U   & -\hat\omega & 0 \\
 0 & 0 & -\overline{\hat\omega} & 1 + \hat U  & -\hat\omega \\
 0 & 0 & 0 & -\overline{\hat\omega} & 1 
\end{array} 
\end{array}}
 \right) 
\end{equation}
\begin{equation}
\hat\mathbb{A}^{\hat A}=\left( {\begin{array}{c|c}
\mathbf{0} & \mathbf{0}\\ \hline 
\mathbf{0} &  \begin{array}{rrrrr}
\hat X^A  & -\hat\xi^A & 0 & 0 & 0 \\
 -\overline{\hat\xi^A} &\hat X^A  & -\hat\xi^A & 0 & 0 \\
 0 & -\overline{\hat\xi^A} & \hat X^A  & -\hat\xi^A & 0 \\
 0 & 0 & -\overline{\hat\xi^A} & \hat X^A & -\hat\xi^A  \\
 0 & 0 & 0 & -\overline{\hat\xi^A} & 0
\end{array} 
\end{array}}
 \right)\,, 
\end{equation}
moreover, the determinant of $\hat\mathbb{A}^{\hat
  r}$  can also be seen to take the form 
\begin{equation}
det\left(\hat\mathbb{A}^{\hat r}\right)=
(\hat U-\hat \omega\overline{\hat \omega}) (1+\hat U)\left[(\hat
  U+1)^2-2\hat 
  \omega\overline{\hat \omega}\right]\,. 
\end{equation}

In proceeding, notice first that since the horizon is non-degenerate,
$\kappa_{\circ}\not=0$, as well as, we have that $\hat {\rm D}(\hat
U-\hat \omega\overline{\hat \omega})|_{{\widetilde{\mathcal{Z}}}_{\hat
u}}=\kappa_{\circ}$, the term $\hat  U-\hat \omega\overline{\hat
\omega}$ must change sign at ${\widetilde{\mathcal{Z}}}_{\hat
u}$. Thereby the character of (\ref{eqV3h}) could, in principle, be
different  on the two sides of the event horizon, $\mathcal{N}$. For
instance, $\hat U-\hat \omega\overline{\hat \omega}$ can be seen to be
positive on the black hole region side of ${\widetilde{\mathcal{N}}}$
which, in accordance with Theorem\,\ref{IR}, implies that  the system
(\ref{eqV3h}), as it stands, is already symmetric hyperbolic on
${\widetilde{\mathcal{N}}}_{\hat u}^{T}$, at least in a small
neighbourhood of ${\widetilde{\mathcal{Z}}}_{\hat u}$, in
$J^+[{\widetilde{\mathcal{Z}}}_{\hat
u}]\setminus{{\widetilde{\mathcal{Z}}}_{\hat  u}}$.  On the other
hand, by multiplying, e.g., the relevant form of $($FR.$4)$, i.e., the
metric equation for $\hat U$, in (\ref{eqV3h}) by $-1$, the yielded
set of PDEs can be seen to possess the form of a first order symmetric
hyperbolic system everywhere  on the domain of outer communication
side except on ${\widetilde{\mathcal{Z}}}_{\hat u}$.  Nevertheless, in
the smooth case the justification of the uniqueness of solutions to
the associated Cauchy problem---which is desired to be done on the
domain of outer communication side---will not be immediately obvious
since the system (\ref{eqV3h}) is irregular. More precisely, in
verifying the uniqueness of the solutions to our specific Cauchy
problem we need to find suitable results guaranteeing the existence
and uniqueness of solutions to first order quasilinear systems with a
`singular' forcing term. Although very few generic results are known
concerning this type PDEs the one we have is apparently very close to
the form of Fuchsian-type equations which have been  studied
extensively in general relativity mainly under the leadership of
Rendall (for related results see, e.g.,
\cite{CN,rendall3,rendall1,rendall2}).  It is worth, however,
emphasising that neither of the available results seems to apply
immediately to the particular form of the PDEs we have. Therefore,  in
the smooth case the problem remains to be open, nevertheless, what has
been discussed above can be summarised as.
\begin{proposition} 
Consider a vacuum spacetime $(M,g_{ab})$ of type B and with a
non-degenerate (future) event horizon ${\mathcal{N}}$. Then the
spacetime metric $g_{ab}$ is uniquely determined, in a neighbourhood
of ${\mathcal{N}}$---besides the black hole region also on the domain of 
outer communication side---by the $2$-metric of the space of Killing
orbits on ${\mathcal{N}}$, if the existence and uniqueness of
solutions to the Cauchy problem relevant for 
(\ref{eqV3h}) can be guaranteed on ${{\mathcal{N}}}_{\hat
u}^T\cap {\mathcal{D}}_{\mathcal{N}}$.
\end{proposition} 

\medskip

Clearly, the Cauchy problem for (\ref{eqV3h}) cannot have unique
solutions in the smooth case unless the uniqueness is guaranteed in
the analytic setting. The rest of this subsection is to 
show that (\ref{eqV3h}) possesses unique solutions in the 
analytic case. 

Start by inspecting (\ref{eqV3h}). It gets to be immediately
transparent that the equation which can be deduced from (NP.6.12e) and
which takes the form
\begin{equation}\label{DhPsi0}
\hat U\cdot\hat  \mathrm{D}\hat\Psi_0 +  \hat X ^A \partial_{\hat x^A}
  \hat\Psi_0-\hat\delta\hat\Psi_1 = (4\hat\gamma-\hat\mu)\hat\Psi_0
  -2(2\hat\tau+\hat\beta)\hat\Psi_1 +3\hat\sigma\hat\Psi_2\,
\end{equation}
is at the centre of the problems. This equation is regular everywhere
on ${\widetilde{\mathcal{N}}}_{\hat u}^{T}$ except at the horizon,
${\widetilde{\mathcal{N}}}$, where $\hat U$ gets to be
zero. Nevertheless, the various order of $\hat\mathrm{D}$-derivatives
of $\hat\Psi_0$, along with that of all the other components in
$\hat\mathbb{V}$, can be determined, by an inductive algorithm, in
terms of $\Psi_2$ and its various inner derivatives on
${\widetilde{\mathcal{Z}}}_{\hat u}$ as follows.

Recall first that, in virtue of Table\,\ref{table:data} and
(\ref{trans}), all the initial values of the components of
$\hat\mathbb{V}$ can be determined in terms of $\Psi_2$ and its
various inner  derivatives on ${\widetilde{\mathcal{Z}}}_{\hat u}$
Notice also that the $\hat\mathrm{D}$-derivative of (\ref{DhPsi0}), in
consequence of the vanishing of $\hat U,\hat X^A,\hat\mathrm{D}\hat
X^A,\hat \mu,\hat \tau$ and $\hat \sigma$ on
${\widetilde{\mathcal{Z}}}_{\hat u}$, reduces to
\begin{equation}\label{DDhPsi0}
\hat\mathrm{D}\hat U\cdot\hat\mathrm{D}\hat\Psi_0 -\hat
\mathrm{D}(\hat\delta\hat\Psi_1) = (4\hat\mathrm{D}\hat\gamma-
\hat\mathrm{D}\hat\mu)\cdot\hat\Psi_0
+(4\hat\gamma-\hat\mu)\cdot\hat\mathrm{D}\hat\Psi_0   
  -2(2\hat\mathrm{D}\hat\tau+\hat\mathrm{D}\hat\beta)\hat\Psi_1
  -2\hat\beta\cdot\hat\mathrm{D}\hat\Psi_1
  +3\hat\mathrm{D}\hat\sigma\cdot\hat\Psi_2 
\end{equation}
on ${\widetilde{\mathcal{Z}}}_{\hat u}$. This relation, along with
(NP.6.8), (NP.6.10d), (NP.6.11b), (NP.6.11c), (NP.6.11d) and
(NP.6.11h),  implies then that
\begin{equation}\label{DDhPsi02}
3\kappa_\circ\cdot\hat\mathrm{D}\hat\Psi_0=
[\hat\delta(\hat\mathrm{D}\hat\Psi_1)
-2\hat\beta\cdot(\hat\mathrm{D}\hat\Psi_1)]
-\hat\rho[\hat\delta\hat\Psi_1 
-2\hat\beta\cdot\hat\Psi_1]
+6[\hat\Psi_0\hat\Psi_2-\hat\Psi_1^2] 
\end{equation}
also holds on ${\widetilde{\mathcal{Z}}}_{\hat u}$. This relation, in
virtue of (NP.6.12a), Table\,\ref{table:data} and (\ref{trans}), along
with the application of the ``edth''-operator several times, justifies 
finally that
\begin{equation} 
\hat\mathrm{D}\hat\Psi_0|_{{\widetilde{\mathcal{Z}}}_{\hat u}}=
\frac{1}{3\kappa_\circ^3}  \left[\frac12\eth\overline{\eth}\eth^2\Psi_2
-10\cdot(\eth\Psi_2)^2 -2\Psi_2\cdot\eth^2\Psi_2 \right]\,.
\end{equation}
By making use of the ``hatted'' form of the Newman-Penrose equations
all the first order $\hat\mathrm{D}$-derivative of the other
components of $\hat\mathbb{V}$ can also be evaluated in terms of $\Psi_2$
and its various inner derivatives on ${\widetilde{\mathcal{Z}}}_{\hat
u}$.

We can, now, proceed inductively to show that all the higher order
$\hat\mathrm{D}$-derivatives of the components of $\hat\mathbb{V}$ can
also be evaluated at ${\widetilde{\mathcal{Z}}}_{\hat u}$ in terms of
$\Psi_2$ and its suitable order of inner derivatives on
${\widetilde{\mathcal{Z}}}_{\hat u}$. Remember that these quantities
are uniquely determined by the specification of the $2$-metric of the
bifurcation surface ${{\mathcal{Z}}}$ given in terms of $\xi^A$, i.e.,
by the only freely specifiable part of the reduced initial data
there. Suppose now, as our inductive assumption, that this can be done
up to the $(p-1)^{th}$-order $\hat\mathrm{D}$-derivatives.  To justify
then the regular determinacy of the $p^{th}$-order
$\hat\mathrm{D}$-derivatives of all the variables in $\hat\mathbb{V}$
except $\hat\Psi_0$ on ${\widetilde{\mathcal{Z}}}_{\hat u}$---this
subset will be denoted as
$\hat\mathrm{D}^p\hat\mathbb{V}_{\hat\Psi_0}$---the $(p-1)^{th}$-order
$\hat\mathrm{D}$-derivatives of ``hatted'' form of the pertinent
Newman-Penrose equations can be used. The key point in our argument is
that, whenever $\kappa_\circ\not=0$, the $p^{th}$-order
$\hat\mathrm{D}$-derivative of (\ref{DhPsi0}) can be used to determine
$\hat\mathrm{D}^p\hat\Psi_0|_{{\widetilde{\mathcal{Z}}}_{\hat  u}}$ as
\begin{equation}\label{DDhPsi023} 
3\kappa_\circ\left(\hat\mathrm{D}^p\hat\Psi_0\right)
    |_{{\widetilde{\mathcal{Z}}}_{\hat 
    u}}=\mathcal{F}\left(\hat\mathbb{V},\hat\mathrm{D}\hat\mathbb{V}, 
\dots, \hat\mathrm{D}^{p-1}\hat\mathbb{V},     
\hat\mathrm{D}^p\hat\mathbb{V}_{\hat\Psi_0}\right)\,,  
\end{equation}
where $\mathcal{F}$ is a sufficiently regular function of its
indicated variables and, according to our inductive hypothesis, all
the indicated lower order $\hat\mathrm{D}$-derivatives,
$\hat\mathrm{D}^q\hat\mathbb{V}$ with $1\leq q< p$, can be evaluated
in terms of $\Psi_2$ and its suitable order of inner derivatives on
${\widetilde{\mathcal{Z}}}_{\hat u}$.  All the above partial results
can then be summarised as.

\begin{theorem}\label{IRI}
Suppose that  $(M,g_{ab})$ is a vacuum spacetime of type B so that the
(future) event horizon ${\mathcal{N}}$ is non-degenerate. Assume that
the spacetime $(M,g_{ab})$ and the event horizon ${\mathcal{N}}$ are
both analytic. Then, the  spacetime metric $g_{ab}$ is uniquely
determined in a neighbourhood of  ${\mathcal{N}}$ on the domain of
outer communication side  once the  $2$-metric of the space of Killing
orbits on  ${\mathcal{N}}$ is fixed.
\end{theorem} 

Notice that, in virtue of (\ref{DDhPsi02}) and (\ref{DDhPsi023}), the
non-genericness of ${\mathcal{N}}$ is of critical importance in the
above argument.

\section{The electrovac black hole spacetimes}\label{EM}
\renewcommand{\theequation}{\ref{EM}.\arabic{equation}}
\setcounter{equation}{0}

This section is to discuss some of the differences which show up in
case of the electrovac black hole spacetimes with non-zero
cosmological constant. We start by providing the explicit form of the
``reduced field equations'', which similarly to the vacuum case,
consist of some of the Newman-Penrose and Maxwell equations or
suitable linear combinations of pairs of the Newman-Penrose  and
Maxwell equations \cite{newman:penrose}.

\medskip

Recall, first, that electromagnetic field, $F_{ab}$, can very
effectively be represented in the Newman-Penrose formalism
\cite{newman:penrose} by making use of the contractions
\begin{eqnarray} 
&& \phi_0 = F_{ab}l^am^b \\
&& \phi_1 = \frac12\,F_{ab}\,\left(l^an^b +  \overline{m}^a m^b
  \right)\\ 
&& \phi_2 = F_{ab}\overline{m}^an^b\,,  
\end{eqnarray} 
while the energy momentum tensor, as given by  (\ref{max}),  can be
represented via the Ricci spinor components, $\Phi_{ij}$, where the
indices $i,j$ take the values $0,1,2$---for their generic definitions
see (NP.4.3b)---which for the considered electrovac case, (\ref{max}),
can be given as
\begin{equation}
\Phi_{ij}=2\,\phi_i\overline{\phi}_j\,. 
\end{equation} 
By making use of these variables the reduced Einstein's equations,
relevant for the electrovac case with cosmological constant,
$\widetilde\Lambda=-6\Lambda$, read as 
\renewcommand{\theequation}{EM.\arabic{equation}}
\setcounter{equation}{0}
\begin{eqnarray}  
&& \mathrm{D}\,\xi ^A={\rho }\,\xi ^A+{\sigma }\,\bar\xi^A \\ 
&& \mathrm{D}\,\omega={\rho\, \omega }+\sigma \,\overline{\omega
  }-\tau  \\  
&& \mathrm{D}X^A={\tau }\,\bar\xi^A+{\bar{\tau}\,\xi }^A \\ 
&& \mathrm{D}\,U ={\tau }\,\overline{\omega }+{\overline{\tau}\,\omega 
    -(\gamma +}  
\overline{\gamma }) \\ 
&& \mathrm{D}\,{\rho} = {\rho }^2+{\sigma \,\overline{\sigma}}
  + \Phi_{00} \\
&& \mathrm{D}\,{\sigma } = 2{\rho \,\sigma }+{{\Psi }_0}  \\  
&& \mathrm{D}\,{\tau } = {\tau
\,\rho }+{\overline{\tau}\,\sigma }+{{\Psi }_1} + \Phi_{01}  \\  
&& \mathrm{D}\,{\alpha}
= {\rho \,\alpha }+{\beta \,\overline{\sigma}} + \Phi_{10}  \\  
&& \mathrm{D}\,\beta = {\alpha \,\sigma }+{\rho \,\beta }+{{\Psi }_1} 
\\  
&& \mathrm{D}\,{\gamma} = {\tau \,\alpha }+{\overline{\tau}\,\beta
}+{{\Psi}_2- \Lambda + \Phi_{11}} \\  
&& \mathrm{D}\,{\lambda} = {\rho\, \lambda
}+{\overline{\sigma}\,\mu + \Phi_{20}} \\  
&& \mathrm{D}\,{\mu } = {\rho \,\mu 
}+{\sigma\,\lambda }+{{\Psi }_2 + 2\,{\Lambda}}\\  
&& \mathrm{D}\,{\nu} =
{\overline{\tau}\,\mu}+{\tau\,\lambda}+{{\Psi}_3 + \Phi_{21}} \\  
&&\hskip-1.5cm  \Delta\, \Psi_0 -\delta\,(\Psi
_1{+{\Phi}_{01}})+\mathrm{D}\, {{\Phi 
  }_{02}} =(4\,\gamma -\mu )\,\Psi _0-2\,(2\,\tau+\beta
)\,\Psi_1+3\,\sigma\,\Psi_2 
\\ && \hskip0cm 
-{\overline{\lambda}}\,{\Phi }_{00}-2\,\beta\,{\Phi }_{01} 
+2\,\sigma\,{\Phi}_{11}+\rho\,{\Phi }_{02}\nonumber\\
&&\hskip-1.5cm   
\Delta\, (\Psi _1-{{\Phi }_{01}}) + \mathrm{D}\,(\Psi _1{-{\Phi
  }_{01}})-\delta\,( \Psi _2+2\,{\Lambda }) +\delta\, {\Phi}_{00}
-\overline{\delta }\,\Psi _0 +\overline{ 
\delta }\,{{\Phi }_{02}}= \\
&& \hskip0cm +(\nu-4\,\alpha)\,\Psi_0+2\,(\gamma+2\,
\rho-\mu)\,\Psi_1  -3\,\tau \,\Psi _2+\,2\,\sigma\,\Psi_3 \nonumber\\ 
&& \hskip+0cm
   +(2\,\tau-\overline{\nu})\,{\Phi
   }_{00}+2\,(\overline{\mu}-\gamma-\rho)\,{\Phi}_{01} -2\,\sigma
   \,{\Phi 
   }_{10} +2\,\tau\, {\Phi }_{11}+(3\,\alpha 
-\overline{\beta })\,{{\Phi }_{02} -2\,\rho\,{\Phi }_{12}} 
\nonumber \\
&&\hskip-1.5cm  \Delta\,(\Psi_2+2\,\Lambda)+ \mathrm{D}\,(\Psi
	  {_2+2\,\Lambda )}-\delta\,(\Psi 
_3+{{\Phi}_{21}})-\overline{\delta }\,(\Psi _1+{{\Phi }_{01}} 
)+\Delta \,{{\Phi }_{00}}+\mathrm{D}\,{\Phi }_{22}= \\
&& \hskip+0cm -\lambda\, \Psi _0+2\,(\nu-\alpha )\,\Psi
	  _1+3\,(\rho-\mu)\, \Psi _2  
-2\,{\overline{\alpha }}\,\Psi _3+\sigma\, \Psi_4 \nonumber \\ && 
+(2\,\gamma +2\,\overline{ 
\gamma }-\overline{\mu })\,{{\Phi }_{00}} {-2\,(\alpha
  +}\overline{\tau })\,{{\Phi }_{01}}  
-{2}\,\tau\,{{\Phi }_{10}}+2\,(\rho-\mu)\,{\Phi }_{11} \nonumber\\
&&-\overline{\lambda }\,{\Phi }_{20}  
+\overline{\sigma }\,{{\Phi }_{02}}+{2\,\beta\,{\Phi
  }_{21}}+\rho\,{\Phi }_{22}\nonumber \\  
&&\hskip-1.5cm  \Delta\, (\Psi _3{-{\Phi }_{21}}) + \mathrm{D}\,(\Psi 
_3{-{\Phi }_{21}})-\delta\,\Psi_4-\overline{\delta }\,(\Psi
_2+2\,\Lambda )  +\delta\,{\Phi }_{20} +\overline{\delta }\,{{\Phi } 
_{22}} = \\
&& \hskip+0cm
-2\,\lambda\, \Psi _1+3\,\nu\, \Psi _2-2\,(\gamma +2\,\mu -\rho)\,\Psi
_3+(4\,\beta -\tau )\,\Psi _4 \nonumber\\ &&  
+2\,\mu \,{{\Phi }}_{10}-{(2\,\beta }-2\,\overline{\alpha
}+\nu)\,{{\Phi }_{20}}-2\,\nu \,{{\Phi }_{11}}  
+{2\,\lambda\, {\Phi }_{12}+2\,(\gamma + 
\overline{\mu }-\rho)\,{{\Phi }_{21}-}}\overline{\tau
}\,{{{\Phi}_{22}}}\nonumber \\  
&&\hskip-1.5cm \mathrm{D}\,\Psi_4-\overline{\delta}\,(\Psi _3+{{\Phi 
  }_{21}})+\Delta\,{{  
\Phi }_{20}} =
-3\,\lambda\, \Psi _2+2\,\alpha\, \Psi _3+\rho\, \Psi _4 \\
&&+2\,\nu\, {{\Phi }  
}_{10}-2\,\lambda\, {\Phi }_{11}   
-{(}2\,\gamma -2\,\overline{\gamma }+\overline{\mu })\,{{\Phi }_{20}} 
{-2\,(\overline{\tau }-\alpha )\,{\Phi }_{21}+}\overline{\sigma
}\,{{\Phi }_{22}}  \nonumber 
\end{eqnarray} 
These equations have to be augmented by the ``reduced Maxwell'' equations
which read as 
\begin{eqnarray} 
&& \Delta \phi_0 - \delta \phi_1 = (2\,\gamma - \mu)\,\phi_0 -
2\,\tau\,\phi_1 + \sigma\,\phi_2\\
&& \Delta \phi_1+ \mathrm{D}\phi_1 - \delta \phi_2 - \overline{\delta}
\phi_0=  (\nu- 2\,\alpha)\,\phi_0 + 2\,(\rho - \mu)\,\phi_1 
-(\tau - 2\,\beta)\,\phi_2 \\
&& \mathrm{D} \phi_2 - \overline{\delta} \phi_1 = -\lambda\,\phi_0 +
\rho\,\phi_2\,.  
\end{eqnarray} 
\renewcommand{\theequation}{7.\arabic{equation}}
\setcounter{equation}{4}{}
It can be proved, completely parallel to the argument outlined in
Subsection\,\ref{nchfr} for the vacuum case, that (EM1) - (EM21) comprise a
determined system for the ``$21$-dimensional'' vector valued variable    
\begin{equation}\label{VEM} 
\mathbb{V}_{_{EM}}=(\xi^A,\omega,X^A,U;\rho,\sigma,\tau,\alpha,\beta,
\gamma,\lambda,\mu,\nu;\Psi_0,\Psi_1, \Psi_2,\Psi_3,\Psi_4; \phi_0,
\phi_1, \phi_2)\,. 
\end{equation}
These equations can also be shown to be ``as good as'' the full set of
the Newman-Penrose and Maxwell
equations. Moreover, when written out these equations in Gaussian
null coordinates $(u,r,x^3,x^4)$ they possess the form of a first order
quasilinear symmetric hyperbolic system, i.e., it can be justified
that they read as
\begin{equation}\label{eqVEM}
(\mathbb{A}_{_{EM}})^\mu \cdot \partial_\mu \mathbb{V}_{_{EM}} +
  \mathbb{B}_{_{EM}}=0, 
\end{equation}
where the matrices $(\mathbb{A}_{_{EM}})^\mu$ and $\mathbb{B}_{_{EM}}$
smoothly depend 
on $\mathbb{V}_{_{EM}}$ and 
$\overline{\mathbb{V}}_{_{EM}}$, moreover, the
matrices $(\mathbb{A}_{_{EM}})^\mu$ are Hermitian and the combination
$(\mathbb{A}_{_{EM}})^\mu(n_\mu+l_\mu)$ is positive definite.

\medskip

In addition, it can also be shown that the reduced set of initial
data includes, besides the usual data relevant for the vacuum
configurations, the value of the cosmological constant,
$\widetilde\Lambda=-6\Lambda$, and the specification of the Maxwell
spinor components $\phi_2$ on ${{\widetilde{\mathcal{H}}_1}}$,
$\phi_0$ on ${{\widetilde{\mathcal{H}}_2}}$ and $\phi_1$ at the
bifurcation surface, $\mathcal{Z}$, 
i.e. it is given as
\begin{equation} 
({\mathbb{V}_{_{EM}}})^{red}_{0}=\{\rho,\sigma,\mu,\lambda,\tau\,;
\,\xi^A; \phi_1 \}|_{\widetilde{\mathcal{Z}}} \cup
\{\Psi_4; \phi_2\}|_{{\widetilde{\mathcal{H}}_1}} \cup 
\{\Psi_0; \phi_0\}|_{{\widetilde{\mathcal{H}}_2}} \cup \{
\Lambda\in\mathbb{R}  
\}\,. 
\end{equation}
It is straightforward to justify that all the results of
Section\,\ref{npfr} generalise to the electrovac
case, whence, by summarising all the above claims, the following
statement can be shown to be true.

\begin{theorem} 
In the characteristic initial value problem to any `reduced initial
data set', $({\mathbb{V}_{_{EM}}})^{red}_{0}$, on ${{\mathcal{H}}_1}
\cup {{\mathcal{H}}_2}$, there always exists a unique solution,
${\mathbb{V}_{_{EM}}}$, everywhere in the domain of dependence of
${{\mathcal{H}}_1} \cup {{\mathcal{H}}_2}$ to the electrovac
Einstein-Maxwell equations.
\end{theorem} 

The determination of a full initial data set
$({\mathbb{V}_{_{EM}}})_{0}$ on ${\widetilde{\mathcal{H}}_1} \cup
{\widetilde{\mathcal{H}}_2}$ can be done completely parallel to the
construction applied in Subsection\,\ref{app}. The relevant results
are collected in   Table\,\ref{table:data2} below.
\begin{table}[h!]
\centering 
\begin{tabular}{|c|c|c|} 
\hline $\phantom{\frac{\frac12}{\frac12}{4}}$ $
 {\widetilde{\mathcal{H}}_1}$ &  ${\widetilde{\mathcal{Z}}}$ &
 ${\widetilde{\mathcal{H}}_2}$ \\ \hline \hline
 $\phantom{\frac{\frac12}{\frac12}{4}}$ $\rho = -u \cdot (\Psi_2 +2
 \Lambda)$ & 
 $\rho =0$ & $\rho =0$ \\  \hline
 $\phantom{\frac{\frac12}{\frac12}{4}}$  $\mu = 0$ & $\mu =0$ & $\mu =
 r \cdot (\Psi_2 +2 \Lambda)$ \\ \hline
 $\phantom{\frac{\frac12}{\frac12}{\frac12}{4}}$
 $\sigma=\lambda=\tau=0$ & $\sigma=\lambda=\tau=0$ &
 $\sigma=\lambda=\tau=0$ \\ \hline
 $\phantom{\frac{\frac12}{\frac12}{4}}$ $\Delta \alpha=\Delta \beta=0$
 & $\alpha,\beta:\ \tau=\overline{\alpha}+\beta=0$  &
 $\mathrm{D}\alpha=\mathrm{D}\beta=0$ \\ \hline
 $\phantom{\frac{\frac12}{\frac12}{4}}$ $\Delta\Psi_2=0$ &
 $\xi^A,\phi_1,\Lambda\ \& 
 \ \alpha,\beta \rightarrow \Psi_2$ & $\mathrm{D}
 \Psi_2 =0$ \\ \hline
$\phantom{\frac{\frac12}{\frac12}{4}}$ $\phi_0 =u\cdot \eth \phi_1$ &
 $\phi_0 =0$ &  $\phi_0 =0$ \\ \hline 
$\phantom{\frac{\frac12}{\frac12}{4}}$ $\Delta \phi_1 =0$ & $\phi_1$ &
 $\mathrm{D}\phi_1 =0$ \\ \hline 
$\phantom{\frac{\frac12}{\frac12}{4}}$ $\phi_2 =0$ &  $\phi_2 =0$ &
 $\phi_2 =r\cdot \overline{\eth} \phi_1$ \\ \hline 
$\phantom{\frac{\frac12}{\frac12}{4}}$
 $\Psi_0=\frac{1}{2}u^2\,\eth^2\Psi_2$
 & $\Psi_0=0$ & $\Psi_0=0$ \\ \hline
 $\phantom{\frac{\frac12}{\frac12}{4}}$  $\Psi_1=u\cdot\eth\Psi_2$ &
 $\Psi_1=0$ & $\Psi_1=0$ \\ \hline
 $\phantom{\frac{\frac12}{\frac12}{4}}$  $\Psi_3=0$ & $\Psi_3=0$ &
 $\Psi_3=r\cdot\overline{\eth}\Psi_2$ \\ \hline
 $\phantom{\frac{\frac12}{\frac12}{4}}$ $\Psi_4=0$ & $\Psi_4=0$ &
 $\Psi_4=\frac{1}{2}r^2\,\overline{\eth}{}^2\Psi_2$ \\ \hline
 $\phantom{\frac{\frac12}{\frac12}{4}}$  (gauge) $\nu=0$\ \ \
 $\rightarrow$ & $\nu=0$ \ \ \ $\rightarrow$ & \ \
 $\nu=\frac{1}{2}r^2\cdot\left(\overline{\eth}\Psi_2
 +2\,\overline{\eth}\phi_1\cdot \overline{\phi_1}\right)$ \ \ \\ \hline
 $\phantom{\frac{\frac12}{\frac12}{4}}$  (gauge) $\gamma=0$\ \ \
 $\rightarrow$ & $\gamma=0$ \ \ \ $\rightarrow$ & \ \
 $\gamma=r \cdot (\Psi_2-\Lambda+\Phi_{11})$  \ \   \\ \hline
\end{tabular}
\caption{\small The full initial data set $({\mathbb{V}_{_{EM}}})_{0}$
relevant for a stationary electromagnetic black hole spacetime, on the 
intersecting null hypersurfaces ${\widetilde{\mathcal{H}}_1} \cup
  {\widetilde{\mathcal{H}}_2}$. }\label{table:data2}
\end{table}

\bigskip

In the electrovac case most of the results  which have been derived
previously only for the vacuum case, remain valid with some slight
modifications. For instance, the counterpart of Theorem\,\ref{IR} can
be formulated as.
\begin{theorem}\label{IR2}
Consider a spacetime $(M,g_{ab})$ of type B and with a non-degenerate
(future) event horizon ${\mathcal{N}}$. Then, both the spacetime
metric $g_{ab}$ and the electromagnetic field are uniquely determined
in the black hole region once the $2$-metric of the space,
${\mathcal{Z}}$, of Killing orbits on ${\mathcal{N}}$, as well as, the
electric potential $\phi_1$ are fixed on ${\mathcal{Z}}$.
\end{theorem} 

Completely parallel to the analysis carried out in
Subsection\,\ref{app}, by making use of the fact that $\mathfrak{K}^a$
is a horizon Killing vector field on $(M,g_{ab})$ and by applying the
null rotation (\ref{nrot}) also to the variables $ \phi_0, \phi_1,
\phi_2$---yielding the transformed ``hatted'' variables given as 
$\hat\phi_0=A^{-1}\phi_0, \hat\phi_1=\phi_1, \hat\phi_2=A\phi_2$---a
system of quasilinear first order PDEs can be derived for the
``hatted'' version of the variables in  $\mathbb{V}_{_{EM}}$. In this
case two of the ``hatted'' version of the field equations, the ones
which can be derived from 
(EM.14) and (EM.19), get to be irregular at the horizon, the latter
possessing the particular form
\begin{equation}\label{Dhphi0}
\hat U\cdot\hat  \mathrm{D}\hat\phi_0 +  \hat X ^A \partial_{\hat x^A}
  \hat\phi_0-\hat\delta\hat\phi_1 = (2\hat\gamma-\hat\mu)\hat\phi_0
  -2\hat\tau\hat\phi_1 +\hat\sigma\hat\phi_2\,.
\end{equation}
  
Nevertheless, the following analog of Theorem\,\ref{IRI} can also be
proved. 
\begin{theorem}
Suppose that  $(M,g_{ab})$ is a spacetime of type B so that the
(future) event horizon ${\mathcal{N}}$ is non-degenerate. Assume that
the spacetime $(M,g_{ab})$ and the event horizon ${\mathcal{N}}$ are
both analytic. Then, both the spacetime metric $g_{ab}$ and the
electromagnetic field are uniquely  determined in a neighbourhood of
${\mathcal{N}}$ on the domain of outer communication side  once the
$2$-metric of the space of Killing orbits on  ${\mathcal{N}}$, as well
as, the electric potential $\phi_1$ are fixed on ${\mathcal{Z}}$.
\end{theorem} 

\medskip

The most significant difference which shows up in the related analysis
is 
due to the presence of a non-zero cosmological constant,
$\widetilde\Lambda=-6\Lambda$. To have a specific example consider
the expansion of the $3$-parameter null geodesic congruence with
tangent pointing towards the direction of the domain of outer
communication, $\mathcal{D}_{\mathcal{N}}$, at ${{\mathcal{N}}}$. In
virtue of Lemma\,\ref{rho} and Table\,\ref{table:data2}, to
guarantee the expansion of this null geodesic congruence to be
non-negative everywhere on ${{\mathcal{N}}}$  the value of the
cosmological constant $\widetilde\Lambda$ (in our
signature) has to be adjusted so that
the inequality $\Psi_2+2\Lambda \leq 0$, or equivalently
$3\,\Psi_2\leq \widetilde\Lambda$, be satisfied throughout 
$\mathcal{N}$.

\section{Final remarks}\label{con}
\setcounter{equation}{0}

In this paper some of the generic properties of stationary distorted
black hole spacetimes were investigated. We would like to emphasise
again  that while all the previous investigations related to distorted
black hole spacetimes were restricted (almost) exclusively to the
static axially symmetric vacuum configurations the geometrical
framework introduced in this paper is  suitable to investigate all the
possible stationary electrovac  distorted black hole solutions.

\medskip

Recall that the event horizon, $\mathcal{N}$, of a generic distorted
black hole spacetime is a ``Killing horizon'' thereby the null
geodesic generators of $\mathcal{N}$ are expansion and shear free.
Accordingly, all of these spacetimes are also spacetimes with an
``{\it isolated horizon}'' the concept of which were introduced (and
evolved) by Ashtekar and his co-workers (see
Refs. \cite{ABL0,ABL,AK,AEB}). It is important to emphasise, however,
that in spite of the great variety of the possible distorted black
hole spacetimes---since they all possess ``the horizon Killing vector
field''---they  form merely a special subclass of spacetimes with an
isolated horizon.

\medskip

Since the spacetimes investigated in this paper are ``black holes'' it
is of obvious interest to know whether the laws of black hole
thermodynamics could be also derived in context of the distorted black
hole spacetimes. In this respect let us mention first that the very
same question had been answered in the confirmatory in case of static
axially symmetric vacuum distorted black holes by Geroch and Hartle
\cite{GH}.  Thereby, it is quite conspicuous to assume that the
analogous investigations can be done---although it was not attempt to
be done in the present paper---in context of generic distorted black
hole spacetimes even though the dimension of the spacetime might be
larger than four or there is a higher variety concerning the topology
of the event horizon (for analogous investigations applicable in case
of $4$-dimensional black hole spacetimes with toroidal or higher genus
horizons see, e.g., \cite{brill}, for relevant higher dimensional
studies see also \cite{Bir}).

As an immediate support of the idea that the laws of black hole
thermodynamics can probably be recovered in case of generic distorted
black hole spacetimes recall first that Lemma 4.2 of the present paper
does, in fact, formulate the content of the relevant ``{\it zeroth
law}''.  In addition, as already mentioned above, the distorted black
hole spacetimes do also fit into the framework of spacetimes with an
isolated horizon within which framework the basic laws of black holes
mechanics had been suitable re-formulated and generalised according to
the needs of the underlying more generic setting (see, e.g.,
\cite{AK,AEB}).  Thereby we expect that the analogous investigations
could also be carried out within the framework of the distorted black
holes  although this should be done so that the usual assumptions
concerning the asymptotic behaviour of the underlying spacetimes are
also weakened suitably.

In this respect let us mention finally that the event horizon of a
distorted black hole is a local causal horizon, in the sense described
by Jacobson and Parentani \cite{JP}. Therefore, one would expect that
a meaningful notion entropy could also be associated with distorted
black holes. This expectation is also supported by the success of
Carlip's proposal which assigns entropy to a black hole by making use
of its  asymptotic near-horizon conformal symmetry
\cite{carlip1,carlip2}. (For some related classical investigations see
also Refs.  \cite{mdv1,mdv2,KLR}.) Since Carlip's approach does not rely
on the global properties of the spacetime it may also be applicable in
case of distorted black holes.

\medskip

Recall that the ``isolated'' asymptotically flat or asymptotically
(locally) anti-de-Sitter stationary electrovac black hole spacetimes
are also included by the set of distorted  black hole spacetimes
investigated in this paper. This set is significantly larger than that
of the isolated black holes since we have not required ``a\,priory''
any sort of asymptotic properties to be possessed by the investigated
distorted black hole spacetimes. Our main result suggests that the
geometry of any $4$-dimensional electrovac distorted black hole is
uniquely determined by the geometry of the bifurcation surface,
along with the specification of the electromagnetic field
there. Thereby, one would expect that by integrating the field
equations outward one should be able to recover the asymptotic region
of the ``isolated'' black hole spacetimes, as well. In this respect it
would be important to find the precise  selection rules
determining the ``isolated'' black hole configurations, in terms of
the data freely specifiable at the bifurcation surface.  Obviously,
the identification of these  conditions would offer more
insight into machinery of the black hole uniqueness argument---which
might be useful to have, especially, in case of higher dimensional
black hole spacetimes---whence the investigation of this issue would
definitely deserve further attention.

\medskip

Concerning the generic distorted black hole spacetimes either of the
following two complementary cases may occur. The null geodesic
congruences transverse to the future event horizon $\mathcal{N}$ of
the black hole are non-contracting everywhere towards the {\it domain
of outer communication}, $\mathcal{D}_\mathcal{N}$, or they might
contract locally. In the former case the cross-sections of
$\mathcal{N}$ may be considered as being ``convex'' everywhere while
in all the other cases they become  locally ``concave''. Whenever a
cross-section is locally concave the associated elementary spacetime
region cannot extend to an asymptotic region.  We would like to
mention that those distorted black hole spacetimes with locally
concave cross-sections may also be considered as being ``rumpled
hairy'' black holes by adopting the concept that a fibre of their hair
is represented by a past directed  null geodesic while the entire hair
of such a black hole is represented by the $(n-1)$-parameter family of
null geodesic congruences starting at the points of the (future) event
horizon $\mathcal{N}$ with past directed null tangent vector field
$-\mathfrak{L}^a$.

According to this picture  even the isolated, i.e.,  asymptotically
flat or asymptotically (locally) anti-de-Sitter, stationary electrovac
black holes---which have been considered for long to have ``no
hair''---get to be hairy, although, their hair is perfectly set.
Notice, however, that the set of distorted electrovac black hole
spacetimes with convex cross-sections has to be larger than that of
the isolated stationary electrovac black holes. This set should
contain, e.g., all of  those configurations for which the null
geodesic intersecting the future event horizon $\mathcal{N}$
transversely are past geodesically complete in the domain of outer
communication. In virtue of the black hole uniqueness results (at
least in the $4$-dimensional case) the corresponding spacetimes cannot
be asymptotically flat but either of the following two cases may
happen. The topology of the cross-sections of the event horizon is
that of a $2$-sphere but the geometry does possess an asymptote
different from that of the isolated configurations or the
cross-sections possess non-spherical topology. In the later case a
large variety of asymptotic structures may occur---the possibility of
which had been noticed long time ago first by Newman and Unti (see the
discussion part of \cite{NU})---since, in principle, there should
exist spacetimes admitting future null infinity with topology
$\scrip\sim \mathbb{R}\times\mathcal{Z}$. The asymptotic symmetries  of
spacetimes with non-spherical sections were investigated by Foster
\cite{fost}, while explicit examples with toroidal sections were
constructed by Schmidt \cite{sch}. Clearly a lot of interesting
related issues have been remained open which would also deserve
further investigations.

\medskip

Let us finally mention that, in virtue of all the above discussions,
it seems to be quite  appropriate to think of the bifurcation surface
of a generic (non-degenerate) $4$-dimensional electrovac distorted
black hole as the unique compact ``carrier'' of the preimage of the
entire associated elementary spacetime region, which can be ``built
up'' by making use of the field equations once the carrier is
provided. In this respect we may also think of the bifurcation surface
of an isolated black hole as a ``holograph'' storing all the
information concerning the associated stationary black hole spacetime.

\section*{Acknowledgements}

This research was supported in part by OTKA grant K67942 and by JSPS
grant L06516.  The author is grateful to Akihiro Ishibashi for helpful
comments and he would also like to thank the Yukawa Institute for
Theoretical Physics for its hospitality.



\begin{thebibliography}{99}


\bibitem{ABL0} A. Ashtekar, C. Beetle and J. Lewandowski: {\it
  Mechanics of Rotating Isolated Horizons},  Phys. Rev. D. {\bf 64}
  044016 (2001)

\bibitem{ABL} A. Ashtekar, C. Beetle and J. Lewandowski: {\it Geometry of
  Generic Isolated Horizons}, Class. and Quant. Grav., {\bf 19},
  1195-1225 (2002) 

\bibitem{AK} A. Ashtekar and B. Krishnan: {\it Isolated and dynamical
  horizons and their applications}, Living Reviews in Relativity, no
  10, 1-77 (2004); 
  http://www.livingreviews.org/lrr-2004-10 

\bibitem{AEB} A. Ashtekar, J. Engle and C.V.D. Broeck:  {\it Quantum
  horizons and black-hole entropy: inclusion 
of distortion and rotation}, Class. Quant. Grav. {\bf 22}, L27-L34 (2005)

\bibitem{Bir} D. Birmingham: {\it Topological black holes in anti-de
  Sitter space}, Class. Quant. Grav. {\bf 16}, 1197-1205 (1999) 

\bibitem{brill} D.R. Brill, J. Louko and P. Peld\'an: {\it Thermodynamics of
  (3+1)-dimensional black holes with toroidal or higher genus
  horizons}, Phys. Rev. D {\bf 56}, 3600-3610  (1997)

\bibitem{bunting}  G.L. Bunting: \textsl{Proof of the uniqueness conjecture
for black holes}, Ph.\,D. Thesis, University of New England, Admirale (1987)

\bibitem{carlip1} S. Carlip, {\it Entropy from conformal field theory
 at Killing horizons}, Class. Quant. Grav. {\bf 16}, 3327 (1999) 

\bibitem{carlip2} S. Carlip: {\it Near-horizon conformal symmetry and
 black hole entropy}, Phys. Rev. Lett. {\bf 88}, 241301 (2002)

\bibitem{carter1}  B. Carter: \textit{Axisymmetric black hole has only two
degrees of freedom}, Phys. Rev. Lett. \textbf{26}, 331-333 (1971)

\bibitem{carter2}  B. Carter: \textit{Black hole equilibrium states, in:
Black Holes}, C. de Witt and B. de Witt (eds.) Gordon and Breach, New York,
London, Paris (1973)

\bibitem{CH} S. Chandrasekhar: {\sl The Mathematical Theory of Black
Holes}, Clarendon Press, Oxford, (1983)

\bibitem{CN} C.M. Claudel and K.P Newman: \textit{The Cauchy problem
  for quasi-linear 
hyperbolic evolution problems with a singularity in the time}, Proc. R. Soc.
London {\bf A 454}, 1073-1107 (1998) 

\bibitem{FK} S. Fairhurst and B. Krishnan: {\it Distorted black holes
  with charge}, Int. J. Mod. Phys. {\bf D 10}, 691-709 (2001)

\bibitem{fost} J. Foster: {\it Asymptotic symmetry and the global
  structure of future null infinity},Int. J. Theor. Phys. {\bf 26},
  1107-1124 (1987) 

\bibitem{friedrich}  H. Friedrich: \textit{On the regular and
  asymptotic characteristic initial value problem for Einstein's
  vacuum field equations}, Proc. Roy. Soc. Lond. A. \textbf{375},
  169-184 (1981) 

\bibitem{friedrich1}  H. Friedrich: \textit{ Cauchy problems for the
  conformal vacuum field equations in general relativity},
  Commun. Math. Phys. \textbf{91}, 445-472 
(1983)  

\bibitem{friedrich2}  H. Friedrich: \textit{ On the hyperbolicity of
  Einstein's and other gauge field equations},
  Commun. Math. Phys. \textbf{100}, 525-543 
(1985) 

\bibitem{frw}  H. Friedrich, I. R\'acz and R.M. Wald: \textit{On the
rigidity theorem for spacetimes with a stationary event horizon or a
compact  Cauchy horizon}, Commun. Math. Phys. \textbf{204}, 691-707
(1999)


\bibitem{FR1} V.P. Frolov and N. Sanchez: {\it Vacuum energy density
  near static distorted black holes}, Phys. Rev. D {\bf
  33}, 1604-1610 (1986)

\bibitem{FR2} A.V. Frolov and V.P. Frolov: {\it Black holes in a
  compactified spacetime}, Phys. Rev. D {\bf 67}, 124025 (2003)

\bibitem{FR3} V.P. Frolov and  A.A. Shoom: {\it Interior of Distorted
  Black Holes}, arXiv:0705.1570 (2007) 


\bibitem{G2}  G.J. Galloway, K. Schleich, D. Witt and E. Woolgar:
\textit{Topological 
censorship and higher genus black holes}, Phys. Rev. D. {\bf 60},
104039 (1999) 

\bibitem{ga1} G.J. Galloway and R. Schoen: {\it A generalisation of
  Hawking's black hole topology theorem to higher dimensions},
  Commun. Math. Phys. {\bf 266}, 571-576 (2006) 

\bibitem{ga2} G.J. Galloway: {\it Rigidity of outer horizons and the
  topology of black holes}, gr-qc/0608118 (2006)

\bibitem{gannon} D. Gannon: \textit{On the topology of spacelike
  hypersurfaces, singularities, and black holes}, Gen. Rel. Grav. {\bf
  7}, 219-232 (1976)

\bibitem{GHP} R. Geroch, A. Held and R. Penrose: {\it A space-time
  calculus based on pairs of null directions}, J. Math. Phys. {\bf
  14}, 874-881 (1973) 

\bibitem{GH} R. Geroch and J. B. Hartle: {\it Distorted black holes},
  J. Math. Phys. {\bf 23}, 680-692 (1982)
 
\bibitem{hawk1}  S.W. Hawking: \textit{Black holes in general relativity},
Commun. Math. Phys. \textbf{25}, 152-166 (1972)

\bibitem{HE}  S.W. Hawking and G.F.R. Ellis: \textsl{The large scale
structure of space-time}, Cambridge University Press (1973)

\bibitem{HO} T. Harmark and N.A. Obers: {\it Black holes on
  cylinders}, JHEP {\bf 05}, 032 (2002) 


\bibitem{HSN} S.A. Hayward, T. Shiromizu, K. Nakao: {\it A
  Cosmological Constant Limits the Size of Black Holes}, Phys.Rev. D
  {\bf 49}, 5080-5085 (1994)


\bibitem{hiw} S. Hollands, A. Ishibashi and R.M. Wald: {A higher
  dimensional stationary rotating black hole must be axisymmetric},
  Commun. Math. Phys. {\bf 271}, 699-722 (2007) 

\bibitem{israel1}  W. Israel: \textit{Event horizons in static vacuum
space-times}, Phys. Rev. \textbf{164}, 1776-1779 (1967)

\bibitem{israel2}  W. Israel: \textit{Event horizons in static electrovac
space-times}, Commun. Math. Phys. \textbf{8}, 245-260 (1968)

\bibitem{israel3} W. Israel and K. A. Khan:  \textit{Collinear
  particles and Bondi dipoles in general relativity}, Nuovo Cimento
  {\bf 33}, 331-344 (1964)

\bibitem{JV} T. Jacobson and S. Venkataramani: {\it Topology of event
  horizons and topological censorship}, 
Class. Quant. Grav. \textbf{12}, 1055-1062 (1995)

\bibitem{JP} T. Jacobson and R. Parentani: {\it Horizon entropy},
  Found. Phys. {\bf 33}, 323-348 (2003) 

\bibitem{rendall3}  S. Kichenassamy and A.D. Rendall: {\it Analytic
description of singularities in Gowdy spacetimes},
Class. Quant. Grav. {\bf 15}, 1339-1355 (1998)

\bibitem{KLR} H.K. Kunduri, J. Lucietti and H.S. Reall: {\it
  Near-horizon symmetries of extremal black holes},
  Class. Quant. Grav. {\bf 24}, 4169­4189 (2007)

\bibitem{szekeres} L.A. Mysak and G. Szekeres: {\it Behaviour of the
  Schwarzschild singularity in superimposed gravitational field},
  Can. J. Phys. {\bf 44}, 617-627 (1966)

\bibitem{mazur}  P.O. Mazur: \textit{Proof of uniqueness of the Kerr-Newman
black hole solutions}, J. Phys. A: Math. Gen. \textbf{15}, 3173-3180 (1982)

\bibitem{mdv1} A.J.M. Medved, D. Martin and M. Visser: {\it Dirty
  black holes: Spacetime geometry and near-horizon symmetries},
  Class. Quant. Grav. {\bf 21}, 3111-3126 (2004) 

\bibitem{mdv2} A.J.M. Medved, D. Martin and M. Visser: {\it  Dirty
  black holes: Symmetries at stationary non-static horizons},
  Phys. Rev. D {\bf 70} 024009 (2004)

\bibitem{ME}  R.C. Myers:  {\it Higher-dimensional black holes in
  compactified space-times}, Phys. Rev. D {\bf 35}, 455-466 (1987)

\bibitem{NU} E.T. Newman and T.W.J. Unti: {\it Behaviour of
  Asymptotically Flat Empty Spaces}, J. Math. Phys. \textbf{3},
  891-901 (1962)   

\bibitem{newman:penrose} E.T. Newman, R. Penrose: \textit{An Approach to 
Gravitational Radiation by a Method of Spin coefficients.}, 
J. Math. Phys. \textbf{3} 566-578 (1962), \textbf{4}, 998 (1963)  

\bibitem{newman66:_note_bondi_metzn_sachs}
E.T. Newman and R. Penrose: {\it Note on the Bondi-Metzner-Sachs
group}, J. Math. Phys., \textbf{7} 863-879 (1966) 

\bibitem{PE} P.C. Peters: {\it Toroidal black holes?},
  J. Math. Phys. {\bf 20}, 1481-1485 (1979)

\bibitem{rw1}  I. R\'{a}cz and R.M. Wald: \textit{Extension of spacetimes
with Killing horizon}, Class. Quant. Grav. \textbf{9}, 2643-2656 (1992)

\bibitem{rw2}  I. R\'{a}cz and R.M. Wald: \textit{Global extensions of
spacetimes describing asymptotic final states of black holes, }Class. Quant.
Grav. \textbf{13}, 539-553 (1996)

\bibitem{r1} I. R\'acz: \textit{On further generalisation of the
rigidity theorem for spacetimes with a stationary event horizon or a
compact Cauchy  horizon}, Class. Quant. Grav. {\bf 17}, 153-178 (2000)

\bibitem{rkill} I. R\'acz: \textit{On the existence of Killing vector 
fields}, Class. Quant. Grav. {\bf 16}, 1695-1703 (1999) 

\bibitem{rkill2} I. R\'acz: \textit{Symmetries of spacetime and
their relation to initial value problems}, Class. Quant. Grav. {\bf
18}, 5103-5113 (2001) 


\bibitem{rendall1} A.D. Rendall: {\it Fuchsian analysis of
  singularities in Gowdy spacetimes beyond analyticity},
  Class. Quant. Grav. {\bf 17}, 3305-3316 (2000)

\bibitem{rendall2}  A.D. Rendall: {\it Fuchsian methods and spacetime
singularities}, Class. Quant. Grav. {\bf 21}, S295-S304 (2004) 


\bibitem{sch} B.G. Schmidt: {\it Vacuum spacetimes with toroidal null
  infinities}, Class. Quant. Grav. {\bf 13}, 2811-2816 (1996)   


\bibitem{Tom} A. Tomimatsu: {\it  Distortion of Schwarzschild-anti-de
  Sitter black holes to black strings}, Phys. Rev. D {\bf 71}, 124044
  (2005) 

\bibitem{wald} R.M. Wald: \textsl{\ General relativity}, University of
Chicago Press, Chicago (1984)


\end{thebibliography}
\end{document}